\documentclass{JHEP3}
\usepackage{amsmath}
\usepackage{axodraw}
\usepackage{epsfig}
\usepackage{amssymb}
\usepackage{mathrsfs}
\usepackage{mathcomp}
\usepackage{cite}
\usepackage[latin1]{inputenc}
\newcommand{\gtaet}{\raisebox{-0.8mm}%
{\hspace{1mm}$\stackrel{>}{\sim}$\hspace{1mm}}}

\def\rap{\ensuremath{Y}}

\keywords{QCD, DIS, Small-$x$, Dipole Model, Saturation}
\preprint{LU-TP 06-47\\
  hep-ph/0702087\\
}

\title{Geometric Scaling and QCD Dynamics in DIS}

\author{Emil Avsar and Gösta Gustafson\\
  Dept.~of Theoretical Physics,
  Sölvegatan 14A, S-223 62  Lund, Sweden\\
  E-mail: \email{Emil.Avsar@thep.lu.se}, \email{Gosta.Gustafson@thep.lu.se}}
  
  \abstract{DIS data from HERA show a striking regularity as 
    $\sigma^{\gamma^* p}$ is a function of the ratio $\tau=Q^2/Q_s^2(x)$ only.
    The scaling function shows a break at $\tau\approx1$, which has been 
    taken as an indication for saturation. However, besides saturation also 
    the transition between dominance of $k_\perp$-ordered (DGLAP) and 
    $k_\perp$-non-ordered (BFKL) evolution contributes to a break around this 
    value of $\tau$, as well as the suppression for small $Q^2$ 
    due to finite quark masses and confinement. 
    In this paper we use a dipole cascade model based on Mueller's dipole model, 
    which also includes energy conservation and pomeron mergings, to investigate
    the contributions of these different effects to the scaling behaviour. 
    As a result we predict that the scaling function for $\tau<1$ 
    will be modified when data for $Q^2>1\,\mathrm{GeV}^2$ become available.
    We also
    investigate the scaling properties of the charm contribution and
    the impact parameter dependence of the saturation scale.}

\begin{document}

\sloppy

\section{Introduction}

Data from deep inelastic scattering (DIS) experiments at small-$x$ 
exhibits an interesting property called geometric scaling \cite{Stasto:2000er}. This means that 
the total $\gamma^*p$ cross section is not a function of the two variables 
$x$ and $Q^2$ separately but rather a function of the combination $Q^2/Q_s^2(x)$ 
only, where the ``saturation scale'' $Q_s$ is defined such that saturation 
is expected to occur at $Q$-values below $Q_s$. 

DIS is quite successfully described  by the Golec-Biernat--W\"usthoff (GBW) model 
\cite{Golec-Biernat:1998js,Golec-Biernat:1999qd}
in which the virtual photon splits into a $q\bar{q}$ dipole long before
the interaction with the proton. The GBW model is also called the saturation model, 
since it explicitely assumes that the dipole-proton cross section, $\sigma_{\mathrm{dp}}$, 
saturates to a constant value $\sigma_0$ as the dipole size, $r$, gets large.
To be more precise, the GBW model assumes that 
\begin{equation}
\sigma_{\mathrm{dp}}=\sigma_0\{1-\exp(-r^2/4R_0^2(x))\},
\label{eq:sigmadp}
\end{equation}
where the ``saturation radius'' $R_0(x)$, identified with $Q_s^{-1}(x)$, 
decreases with decreasing $x$.
In the GBW fit $Q_s^2$ has the form
\begin{equation}
R_0^{-2} =Q_{s,GW}^2 =Q_0^2 \left(\frac{x_0}{x}\right)^\lambda 
\,\,\,\,\mathrm{with}\,\,\, Q_0=1 \mathrm{GeV},\, x_0=3\!\cdot\! 10^{-4},\, \lambda=0.29.
\label{eq:R0}
\end{equation}
We thus see that $\sigma_{\mathrm{dp}}$ is a function of $r/R_0(x)$ only, 
and consequently it satisfies ``geometric scaling''. 
For massless quarks this also implies that the $\gamma^*p$ cross section 
is a scaling function of $\tau = Q^2 R_0^2 = Q^2/Q_{s,GW}^2$, and this
feature is indeed confirmed by experimental data as demonstrated in
ref.~\cite{Stasto:2000er}. (This reference also finds scaling if
the relation $\ln R_0^2\sim \lambda \ln x$ is replaced by 
$\ln R_0^2\sim -x^{-0.08}$.)

Plotting the observed $\gamma^*p$ cross section as 
a function of $\tau $ one can also see that the growth of the 
cross section towards smaller $\tau$
is reduced noticeably when $\tau \lesssim 1$, corresponding to values 
satisfying $Q^2 \lesssim Q_{s,GW}^2$. Later analyses
including DGLAP evolution for larger $Q^2$ \cite{Bartels:2002cj} 
improve the agreement with the HERA data, with the 
consequence that saturation is beginning at somewhat smaller
$\tau$-values. This good agreement together
with the success of the GBW model in describing 
diffractive data, has been taken by some authors as a proof that saturation 
exists, and that it has been observed at HERA.

Including the finite quark mass in the wavefunction describing the
$\gamma^*-q\bar{q}$ 
coupling introduces scale-breaking effects. In particular the 
charm quark gives a large contribution to the cross section, which
is phased out for
$Q^2 \lesssim 4 m_c^2$. In ref.~\cite{Golec-Biernat:1998js} a fit 
including the charm quark mass actually shifts 
the expected onset of saturation to much smaller $x$-values, with 
$x_0= 0.4\!\cdot\! 10^{-4}$ in the expression for $R_0$ or $Q_s$ in
eq.~(\ref{eq:R0}). Such a shift is confirmed in the analysis in
ref.~\cite{Golec-Biernat:2006ba}, which includes the mass of the charm and 
beauty quarks into the DGLAP-improved formalism of ref.~\cite{Bartels:2002cj}.

For small $Q^2$ also the mass of the light quarks becomes important.
In refs.~\cite{Stasto:2000er, Golec-Biernat:1998js} this is
taken into account by replacing the variable $x = Q^2/W^2$
in the definition of $R_0$ or $Q_{s,GW}$ by 
\begin{equation}
\bar{x}=x (1+4m_f^2/Q^2) = \frac{Q^2+4m_f^2}{W^2}.
\label{xbar}
\end{equation}
In this way the scaling relation can
be studied also for very small $Q^2$ and for photoproduction.

The test of scaling in the region of small $\tau$ is, however, limited
by the fact that small values of 
$\tau$ are reached \emph{only} for small $Q^2$,
% typically below 1 GeV$^2$,
since larger $Q^2$-values would need energies not accessible at the 
HERA accelerator. 
Thus the data for $\tau< 0.5-1$, where the change in the slope of the 
$\gamma^*p$ cross section is observed, are all obtained for $Q^2$-values 
smaller than 1 GeV$^2$, which means that they are all in the non-perturbative region.
The limited kinematical range at HERA also implies that there is little
overlap between data at different $Q^2$ for fixed $\tau$, which implies that 
it is relatively easy to achieve a scaling result by adjusting the quark
masses.

The question of scaling for $Q^2 > Q_s^2$ is discussed by Iancu et 
al. in ref.~\cite{Iancu:2002tr}. These authors argue that if $Q_s^2$
is defined as the scale where the scattering probability is of order 1,
then the BFKL evolution equation implies that the
quark and gluon distributions have to obey geometric scaling
in the range $1\lesssim \ln(Q^2/Q_s^2) \ll \ln(Q_s^2/\Lambda_{QCD}^2)$.
This rather wide range in $Q^2$ results from the fast diffusion
in $\ln k_\perp^2$ and the fast growth towards small $x$
in the leading log BFKL evolution. Beyond this range in $Q^2$ the 
BFKL diffusion is gradually replaced by the (not explicitely scaling) double 
leading log result.

A numerical analysis of the diffusion in the BK
equation is presented by Golec-Biernat et al. \cite{Golec-Biernat:2001if},
including non-leading effects from a running coupling and from the
so called kinematical constraint. The result from a running $\alpha_s$
is that $\ln Q_s^2$ for high energies grows $\sim \sqrt{Y}$, rather 
than proportional to $Y$, as assumed in eq.~\eqref{eq:R0}. 
It also significantly reduces the diffusion 
into the region of large $k_\perp^2$. The kinematical constraint reduces
this diffusion further, and these results therefore put a question
mark for the large scaling region dominated by BFKL dynamics obtained
in ref.~\cite{Iancu:2002tr}. A related work is presented by Kwieci\'nski
and Sta\'sto \cite{Kwiecinski:2002ep}, where they study DGLAP evolution starting from 
scaling initial conditions on the line $Q^2=Q_s^2$, with $Q_s^2$ 
determined by some unspecified dynamics. The result from this approach
is also that scaling is approximately preserved in a large domain 
above the line $Q^2=Q_s^2$.

In the 
past few years it has been observed that a scaling feature is inherent in
the asymptotic solutions to the evolution equations in high energy QCD.
It was also realized that the non-linear Balitsky-Kovchegov (BK) 
equation, which is the mean field version of the more general Balitsky-JIMWLK 
hierarchy (B-JIMWLK), is similar to a certain type of equation, well known 
in statistical physics, called the Fisher-Kolmogorov-Petrovsky-Piscounov 
(FKPP) equation, which is known to have traveling wave solutions \cite{Munier:2003vc, Munier:2004xu}. 
Written for a function $u(x,t)$, which depends on $x$ and the time $t$, the solution
for large $t$ has the form of a traveling wave, $u(x-vt)$, where $v$ is the 
speed of the wave. The similarity is expressed by the fact that the 
BK equation lies in the same universality class as the FKPP equation. 

More recently, the importance of fluctuations in small-$x$ evolution has been
better understood \cite{Iancu:2004iy}, and this has led to the modification 
of the original version 
of the B-JIMWLK hierarchy into a new hierarchy of equations, which include both
fluctuations and saturation effects. These new equations are also referred to
as pomeron loop equations, since they contain both pomeron splittings and 
pomeron mergings in the evolution. This hierarchy of equations can also be 
written as a single Langevin equation, which is very similar to
what is called the stochastic-FKPP (s-FKPP) equation \cite{Iancu:2004iy}. The study of 
the asymptotic
behaviour of the solution to this equation leads to the prediction 
of a new type of scaling law, called diffusive scaling \cite{Iancu:2004es}, 
which is expected to hold at very high energies. It ought to be 
emphasized that the traveling wave solutions discussed here
are asymptotic solutions expected to be relevant at extremely high energies,
and therefore cannot be used to explain the scaling observed in the HERA 
energy regime.

%Even though saturation effects lead to geometric scaling, this by itself 
%cannot be taken as a proof of the existence of saturation at HERA. 

The above discussion raises three important questions:
 
i) What is the importance of saturation for the scaling behaviour? 

ii) What is the dynamical
mechanism behind the scaling observed for $\tau>1$, i.e. for $Q^2>Q_s^2$?

iii) Will the cross section 
still be scaling for $\tau\lesssim 1$, when data for larger $Q^2$ are 
available? 

In this paper we will argue that geometric scaling is expected also in
the absence of saturation, and not only in the region dominated
by BFKL diffusion, but also in the double leading log domain, where
$k_\perp$-ordered (DGLAP) evolution chains are most important.
Scaling appears naturally in a dipole cascade model \cite{Avsar:2005iz, Avsar:2006jy}, which is 
based on Mueller's 
dipole evolution \cite{Mueller:1993rr,Mueller:1994jq,Mueller:1994gb} 
but also includes energy-momentum conservation, pomeron
merging, and a simple model for the proton. The MC implementation of the model reproduces both
$F_2$ data from HERA and the total cross section in proton-proton scattering.
This model shows geometric scaling for $Q^2$ below as well as above $Q_{s,GW}^2$,
and for the one pomeron contribution as well as for the full unitarized result.
In the model the transition between BFKL diffusion and $k_\perp$-ordered
evolution occurs for $Q^2$ quite close to $Q_{s,GW}^2$, and there are
three different effects which all contribute to the change in the scaling
curve for $\tau \approx 0.5-1$: Saturation, the BFKL-DGLAP transition,
and the finite effective masses for the light $u$-, $d$-, and $s$-quarks.

In order to get an intuitive understanding of the scaling feature we will
besides the results of the MC simulations also
discuss two simple approximations, which contain the basic features of
DGLAP-BFKL evolution and the colour dipole cascades. 

Diffractive excitation or rapidity gap events correspond to a large fraction
of the events at HERA. Diffractive scattering is related to the 
impact parameter dependence of the interaction, 
and in section \ref{sec:impact} we will discuss
how the speed of the traveling wave varies with impact 
parameter, averaging out to the velocity observed in the experimental data.

The paper is organized as follows. In the next two sections we briefly discuss
the DIS cross section, the concept of geometric scaling, 
and the dipole cascade approach to DIS. In section \ref{sec:saturation}
we discuss the effect of the charm contribution together 
with the effect of saturation on the total cross section. In section
\ref{sec:understanding}, we present the two simple approximations to the 
full model as mentioned above. 
The scaling properties of the charm structure function are studied
in section \ref{sec:charm}, and in section \ref{sec:smallQ} we  
concentrate on the region $Q^2 < Q_s^2$ and study the effects of confinement 
and finite quark masses for small $Q^2$ below 1 GeV$^2$ and 
the scaling properties of the cross section in this region. In section 
\ref{sec:impact} we investigate the behaviour of the scattering amplitude 
for different impact parameters and how the scaling feature
varies with impact parameter. Finally, in section \ref{sec:conclusion}, we
reach at our conclusions. 

\section{DIS and Geometric Scaling}
\label{sec:dis}
In the dipole description of DIS the virtual photon, long before the 
interaction with the proton, splits into a $q\bar{q}$ pair which then 
interacts with the proton. The transverse separation between the quark 
and the antiquark in such a dipole is denoted  $\pmb{r}$, and their
fractions of the $\gamma^*$ longitudinal momentum $z$ and $1-z$. 
The coupling 
of the $\gamma^*$ to the $q\bar{q}$ pair is well known and the leading 
order result reads
\begin{eqnarray}
\vert \psi_L(z,r)\vert^2&=&\frac{6\alpha_{em}}{\pi^2}\sum_f 
e_f^2Q^2z^2(1-z)^2K_0^2\left(\sqrt{z(1-z)Q^2+m_f^2}\,r\right) \nonumber \\
\vert \psi_T(z,r)\vert^2&=&\frac{3\alpha_{em}}{2\pi^2}
\sum_fe_f^2\left\{[z^2+(1-z)^2](z(1-z)Q^2+m_f^2)K_1^2\left(\sqrt{z(1-z)Q^2+m_f^2}\,r\right)\right. \nonumber \\
&+&m_f^2\left.K_0^2\left(\sqrt{z(1-z)Q^2+m_f^2}\,r\right) \right\}.
\label{eq:psigamma}
\end{eqnarray}
Here $\psi_L$ and $\psi_T$ denote the longitudinal and transverse 
wave functions respectively. $K_0$ and $K_1$ are modified Bessel 
functions and the sum $\sum_f$ runs over all active quarks flavours, 
with mass $m_f$ and electric charge $e_f$. The $\gamma^*p$ total 
cross section can then be written as 
\begin{eqnarray}
\sigma_{\gamma^*p}^{tot}=\int d^2\pmb{r}\int_0^1 dz\{\vert \psi_L(z,r)\vert^2
+\vert \psi_T(z,r)\vert^2\} \sigma_{dp}(z,\pmb{r}).
\label{eq:dissigma}
\end{eqnarray}
This cross section is related to the $F_2$ structure function 
via the relation
\begin{eqnarray}
F_2 = \frac{Q^2}{4\pi \alpha_{em}}\sigma_{\gamma^*p}^{tot}.
\label{eq:F2}
\end{eqnarray}

The factor $\sigma_{dp}(z,\pmb{r})$ in \eqref{eq:dissigma} above 
denotes the dipole-proton cross section. In the 
GBW model this cross section is assumed to have the form given in
eq.~\eqref{eq:sigmadp}.
If we only consider the three light quarks and neglect their masses, 
we see that the dipole size can be rescaled $r \rightarrow r/R_0$, which
implies that 
the result only depends on the scaling variable $\tau \equiv Q^2/Q_s^2$.

The charm quark is known to give a significant contribution to the cross section,
and the heavy charm quark must obviously have a large effect on the scaling properties.
This will be discussed further in sections~\ref{sec:saturation} and
\ref{sec:charm}. 
However, also for the light quarks the finite masses
will have significant effects for small $Q^2$, which
will be discussed in section~\ref{sec:smallQ}.

\section{The dipole cascade model for DIS}
\label{sec:cascademodel}

We will in this section shortly describe the model we use to study the effect on 
geometric scaling from different features in the QCD evolution.

DIS at small $x$ is dominated by gluonic cascades due to the $1/z$ singularity 
in the gluon splitting function. 
At large $Q^2$ the cascade can be described by DGLAP evolution, where the 
gluons are strictly ordered in transverse momentum. For smaller $Q^2$ 
also non-ordered gluon chains are important, and the $k_\perp$-ordered
DGLAP evolution is replaced by the $x$-ordered BFKL evolution. The ordering
in $k_\perp$ when $Q^2 \rightarrow \infty$ and the ordering
in energy when $x \rightarrow 0$ follow both from an ordering in angle or,
equivalently, in rapidity.
Such an ordering is a consequence of soft gluon interference, and is the basis
for the CCFM model \cite{Catani:1990yc,Ciafaloni:1988ur}, 
which reproduces DGLAP and BFKL evolution in their
respective domains of applicability with a smooth transition in between.
The CCFM model is reformulated and generalized in the 
Linked Dipole Chain (LDC) \cite{Andersson:1995ju}
model. Here some emissions, which are treated as initial state radiation in 
CCFM,
are instead included as final state radiation, with the result that the
``non-Sudakov'' formfactors disappear, and the cascade becomes symmetric
when exchanging the role of the projectile and the target.
The remaining gluons (called primary gluons in ref.~\cite{Andersson:1995ju} 
and backbone gluons in ref.~\cite{Salam:1999ft}) define totally the 
structure of the final state.

At very high energies the density of gluons becomes very large, and 
nonlinear effects are needed to tame the exponential growth in the above 
linear evolutions, which would otherwise break the unitarity
limit. Saturation and multiple interactions are easier to take into account
in a formulation in transverse coordinate space since at high energies the 
transverse coordinate does not change between repeated subcollisions,
while these collisions do change the transverse momenta. This is exploited in the GBW dipole model
and the dipole cascade model by Mueller 
\cite{Mueller:1993rr,Mueller:1994jq,Mueller:1994gb}. 

The evolution in Mueller's model reproduces the leading order (linear) 
BFKL evolution. It starts from a colour singlet $q\bar{q}$ pair.
The quark and the antiquark emit gluons coherently, forming a colour dipole.
The original dipole is then split in two dipoles formed by the $qg$ and $g\bar{q}$
systems. The new dipoles split repeatedly forming a dipole cascade. In the
leading log approximation the probability
per unit rapidity $Y$ 
for a dipole with transverse coordinates $\pmb{x}$ and $\pmb{y}$ 
to split emitting a gluon at $\pmb{z}$ is given by
(to leading log accuracy $Y\equiv \ln 1/x$ and the true rapidity are equivalent) 
\begin{eqnarray}
\frac{d\mathcal{P}}{d\rap}=\frac{\bar{\alpha}}{2\pi}d^2\pmb{z}
\frac{(\pmb{x}-\pmb{y})^2}{(\pmb{x}-\pmb{z})^2 (\pmb{z}-\pmb{y})^2}
\label{eq:dipolesplit}
\end{eqnarray}

The splitting probability in eq.~(\ref{eq:dipolesplit}) is singular for small
dipole sizes $\pmb{x-z}$ or $\pmb{z-y}$. These singularities have to be 
screened by a cutoff, but the small dipoles have
also a small probability to interact with the target, and therefore
the total cross section is finite when the cutoff goes to zero.
This implies that a lot of non-interacting virtual dipoles are created
in the process, which also makes computer simulations difficult \cite{Salam:1996nb,Salam:1995uy}.

It is well known that a significant part of next to leading corrections
are related to energy-momentum conservation \cite{Salam:1999cn}, and
in refs.~\cite{Orr:1997im, Andersen:2003gs} 
it is demonstrated that energy conservation has a large effect
on the small $x$ evolution. Relating the dipole size $r$ to $1/k_\perp$
there are great similarities between Mueller's model and the LDC model.
In ref. \cite{Avsar:2005iz} these similarities were used to implement 
energy conservation in the dipole cascade formalism. The result is a dynamical
cutoff for small dipoles, and the remaining emissions are ordered in both light 
cone variables $p_+$ and $p_-$, similar to the ordering in the 
LDC model. A dominant fraction of the virtual emissions is here eliminated,
leaving mainly the ``primary gluons'' mentioned above. As a result
the exponential growth for small $x$ is significantly reduced.
It also greatly improves the efficiency of the MC simulation, removing
the difficulties encountered in refs.~\cite{Salam:1996nb,Salam:1995uy}. 

\FIGURE[t]{
  \includegraphics[angle=270,  scale=0.8]{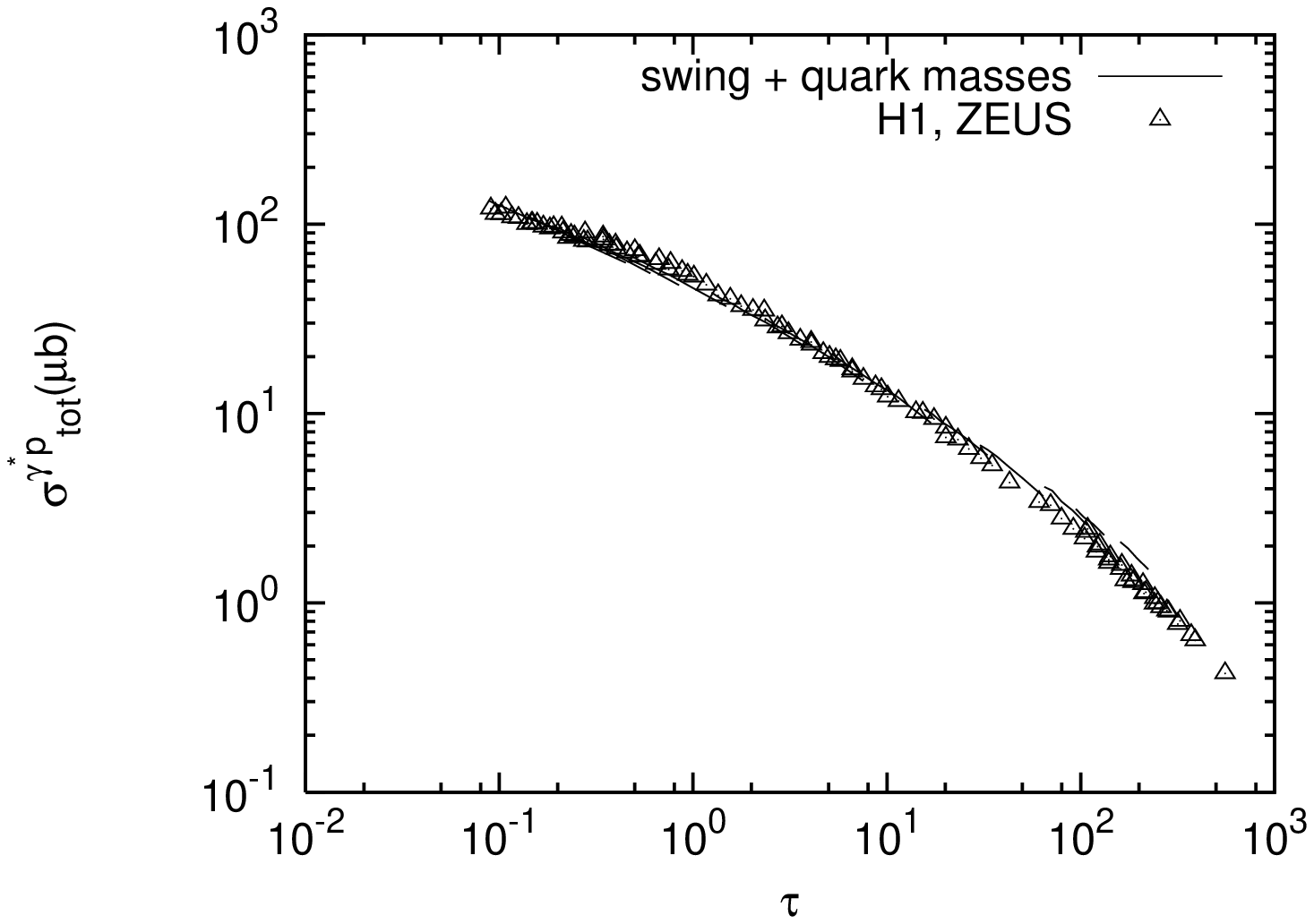}
  \caption{\label{fig:mcscaling}  Our full results for 
    the $\gamma^*p$ total cross section plotted as a 
    function of the scaling variable $\tau=Q^2/Q_{s,GW}^2$. 
    Here saturation effects are included both in the collision between the $\gamma^*$  
    and the proton cascades and within the evolution in each individual cascade,
    via the so called dipole swing.
    We have used a charm mass of 1.4 GeV and an effective light quark mass of 60 MeV 
    for the $u, d$ and $s$ quarks. 
    Data points are taken from \cite{Adloff:2000qk} and \cite{Breitweg:2000yn}.
    Results are presented for $Q^2$ ranging from 0.3 GeV$^2$ to 90 GeV$^2$,
    and with the same kinematics for the data and the model.
}}

Mueller's cascade
includes saturation effects from multiple collisions in the Lorentz
frame chosen for the calculation, but not the effects of pomeron
merging in the cascades. The result is therefore not Lorentz frame
independent. In ref.~\cite{Avsar:2006jy} we improved our model by
allowing (colour suppressed) recouplings of the dipole chain during the 
evolution, a ``dipole swing'', which leads to an almost frame independent 
formalism.
In this paper we also introduced a simple model for the proton in terms
of three dipoles. 

The dipole splitting is calculated in perturbative QCD, and therefore
the model is meant to work in the perturbative regime, which means
not too small $Q^2$. In figure \ref{fig:mcscaling} we show results for 
$\sigma^{\gamma^*p}$ as a function of the scaling variable $\tau 
\equiv Q^2/Q_s^2$ for different $Q^2$ above 0.3 GeV$^2$. The results 
include both the dipole swing and multiple collisions and for
$Q_s(x)$ we here use the definition by Golec-Biernat and W\"usthoff, 
given by \eqref{eq:R0}. For these $Q^2$-values a scale-breaking effect 
is obtained from the charm mass, for which we use the value 1.4 GeV. The
theoretical result is presented for the same kinematical variables
as the experimental data, and we see that there is a very good
agreement between theory and data. For small $Q^2$ the result
is sensitive to confinement effects and effective quark masses.
These problems will be further discussed in sec.~\ref{sec:smallQ}. 

\section{Effects of saturation and the charm contribution}
\label{sec:saturation}

The effects of saturation and of the large charm quark mass are 
illustrated in figure \ref{fig:mcscaling2}. The results correspond to 
$Q^2$ between 0.75 and 90 $\mathrm{GeV}^2$, and therefore the light quark 
masses can be neglected. The dotted lines show the results of the full model,
presented in fig.~\ref{fig:mcscaling}.
The solid lines show the result when the charm mass is set to zero.
The deviation between these curves
therefore shows the effect of the charm quark mass.
Neglecting also the swing gives the long-dashed curves, and finally including
only the single pomeron term in the collisions gives the short-dashed curves.
The difference between these and the solid lines therefore represents
the effect of saturation.

We first note that saturation effects from multiple 
collisions and the swing have a relatively small effect for $\tau>1$, but 
grow for smaller $\tau$, and reduce the cross 
section by approximately a factor 2.5 for $\tau = 0.1$.
The effect of the charm quark mass is, as expected,
also largest for smaller $\tau$. The charm contribution is about 30\% for
$\tau=100$ and 10\% for $\tau=1$, which should be compared with
40\% for zero mass charm quarks.

We note in particular that also the one-pomeron result satisfies 
geometric scaling, even down to $\tau=0.1$.
Thus the scaling feature alone is not enough to prove the existence of 
saturation. The scale-breaking effect from the charm contribution
will be further discussed in sec.~\ref{sec:charm}. The dominance of the 
light quark contributions, and the strong correlation between $Q^2$ and $x$
due to the limited HERA energy, imply that this effect is not so evident in 
the results shown in figs.~\ref{fig:mcscaling} and \ref{fig:mcscaling2},
although we can see a small shift between the curves for different
$Q^2$ in fig.~\ref{fig:mcscaling2}. 

\FIGURE[t]{
  \includegraphics[angle=270,  scale=0.8]{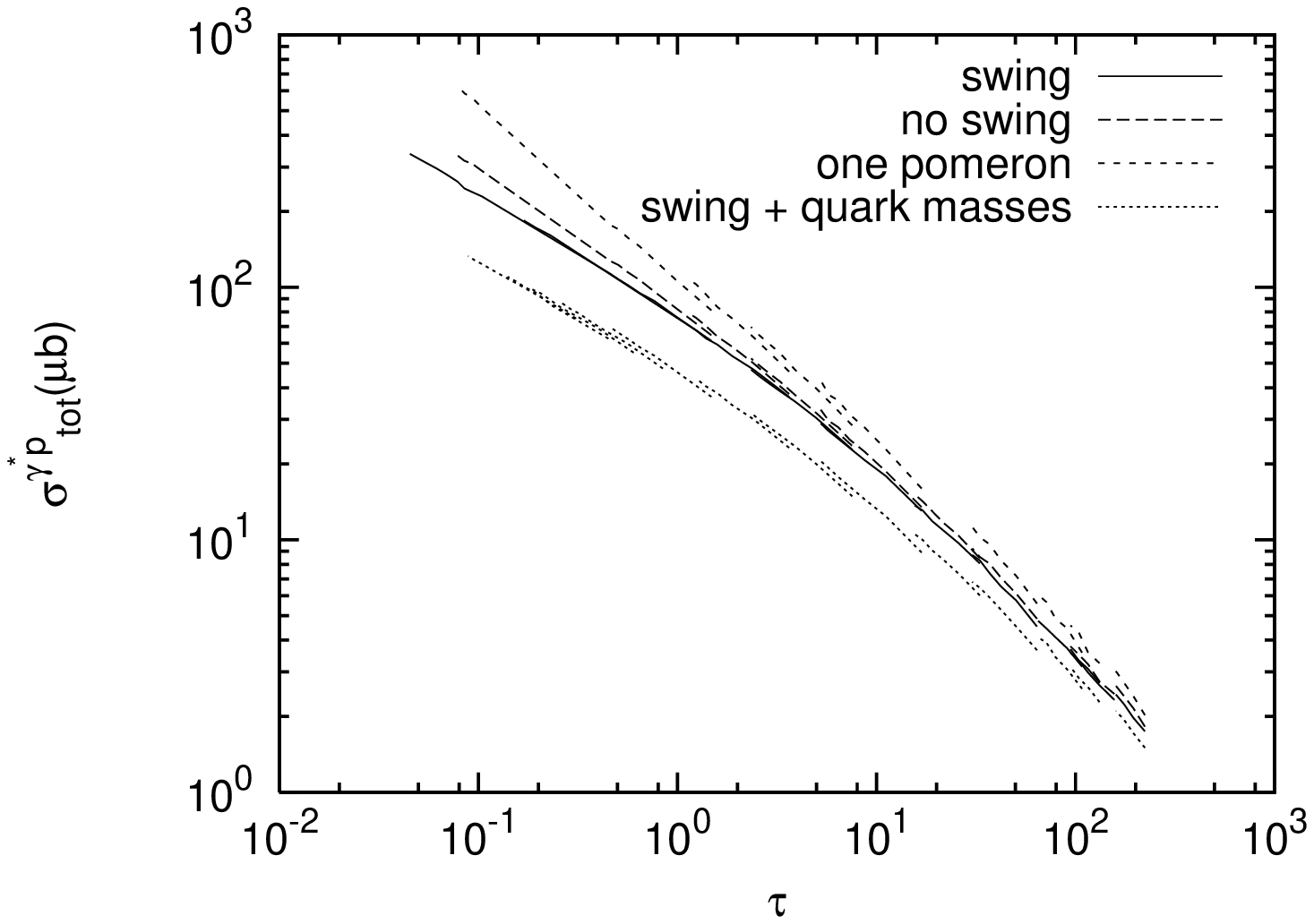}
  \caption{\label{fig:mcscaling2} The top three curves show the $\gamma^*p$ total cross 
    section obtained from our 
    Monte Carlo simulation for $Q^2$ ranging from 0.75GeV$^2$ to 90GeV$^2$ and 
    for zero quark mass for 4 flavours.  
    The solid lines show the results for evolution including the dipole swing and 
    also multiple collisions, the long-dashed lines show the results without the 
    swing but including multiple collisions, while the short-dashed lines are results 
    without the swing and also without multiple collisions. It is seen that all 
    three results satisfy geometric scaling. The bottom curve represents our 
    full results shown in fig \ref{fig:mcscaling}. } }

Although saturation reduces the result for $\tau<0.5$, there is a more 
clear break in the experimental data. We note, however, that
the experimental points, which show this break,
all correspond to small $Q^2$-values below 0.5 GeV$^2$. Here
we expect also non-perturbative effects to be important. 
The wave functions in eq. \eqref{eq:psigamma} extend to
very large transverse separations $r$, and such large dipoles must be
suppressed by confinement effects. 
We also note that energy limitations imply that for fixed 
$Q^2$ the experimental points lie in a rather small $x$-interval, which 
implies that the points for different $Q^2$ have a limited 
overlap. The scaling features for small $Q^2$ will be further
discussed in sec.~\ref{sec:smallQ}.

\section{Understanding geometric scaling in the linear evolution}
\label{sec:understanding}

The saturation model gives a motivation for geometric scaling in the kinematic range
$Q^2 < Q_s^2$, but it does not give a reason for the observed scaling behaviour
for $Q^2 > Q_s^2$ at experimentally feasible energies at current accelerators. 
The arguments in ref.~\cite{Iancu:2002tr} imply that the cross section
should scale as a function of $Q^2/Q_s^2(x)$ within a wide kinematic
region, 
\begin{equation}
1\lesssim \ln(Q^2/Q_s^2)\ll \ln(Q_s^2/\Lambda_{QCD}^2),
\label{eq:Qregion}
\end{equation}
which is dominated 
by BFKL diffusion. Here $Q_s^2(x)$ is defined as the scale
where the scattering probability is of order 1. As mentioned in the
introduction this is a consequence of the fast growth towards small $x$
and wide diffusion to large $k_\perp$ in the leading log BFKL evolution.
Both these effects are strongly damped by non-leading effects 
\cite{Golec-Biernat:2001if}, which therefore reduce the region dominated
by linear BFKL evolution. (We also note that both references
\cite{Golec-Biernat:2001if} and \cite{Iancu:2002tr} point out that
with a running coupling the BFKL evolution leads to a saturation scale
which satisfies $\ln Q_s^2 \sim \sqrt{Y}$ rather than $\ln Q_s^2 \sim Y$.)
For larger $Q^2$, $k_\perp$-ordered evolution 
should be dominant and Kwieci\'nski and Sta\'sto \cite{Kwiecinski:2002ep} 
have assumed that DGLAP evolution is applicable in the whole region
$Q^2>Q_s^2$, with scaling initial conditions at $Q^2=Q_s^2$.  
They then found that although not perfectly scaling, the result showed 
approximate scaling in the same kinematical domain specified by
eq.~\eqref{eq:Qregion}. Their arguments do, however,
not explain the shape of the saturation line. 

We will below argue that scaling is expected both in the region dominated by
BFKL diffusion and in the $k_\perp$-ordered double leading log (DLL) regime 
at larger $Q^2$. 
(In the DLL regime this is not obvious from the analytic expressions, but
follows from a numerical analysis.) Furthermore we find that the transition
between the DLL ($k_\perp$-ordered) and BFKL ($k_\perp$-non-ordered) regimes 
is given by $Q^2 = Q_{\mathrm{limit}}^2 \propto x^{\lambda_{BFKL}}$, 
where $\lambda_{BFKL}$ is the exponent in the solution to the BFKL evolution,
estimated to be around 0.3. This implies that $Q_{\mathrm{limit}}^2$ is very close to
$Q_{s,GW}^2$, leaving only a very small region where the linear BFKL evolution
is dominating.

The qualitative features of the QCD evolution at small $x$ are present already
in the leading log $1/x$ results, and we will in sec.~\ref{sec:scaleLL}
discuss a toy model describing the BFKL-DLL transition \cite{Gustafson:2001iz}. 
Non-leading corrections from \emph{e.g.} energy 
conservation and a running coupling are important for the quantitative result. In 
sec.~\ref{sec:linearcascade} we try to isolate the most important features of
the full dipole cascade MC, to get an intuitive picture of geometric scaling
in the non-saturated region.

\subsection{Leading log approximation}
\label{sec:scaleLL}

When both $1/x$ and $Q^2$ are large 
the gluon distribution is given by
ladders which are ordered in $k_\perp$ and where the splitting
function is dominated by the $1/z$ pole. This corresponds to 
the double log approximation, and for a running coupling 
%(with the definition $\bar{\alpha} \equiv \alpha_0 /\ln (q_\perp^2/\Lambda^2)$)
the gluon density is given by
\begin{equation}
g(x, Q^2)
\sim \exp\left(2\sqrt{\alpha_0 \ln\ln Q^2 \ln 1/x}\right),\,\, 
\mathrm{where}\,\,\bar{\alpha}(Q^2) \equiv \frac{\alpha_0}{\ln (Q^2/\Lambda^2)}.
\label{eq:DLLA}
\end{equation}
%we define the parameter $\alpha_0$ by the relation
%\begin{equation}
%\bar{\alpha} \equiv \frac{\alpha_0}{\ln (q_\perp^2/\Lambda^2)}.
%\label{alfabar}
%\end{equation}
This expression does not scale exactly with $Q_s^2 \propto x^{-\lambda}$. 
Neglecting the very slow variation of $\ln\ln Q^2$ the cross section
$\sim g/Q^2$ scales
with $Q_s^2 \propto \exp(\lambda'\sqrt{\ln 1/x})$, with some parameter $\lambda'$, 
and in ref.~\cite{Gelis:2006bs} it is pointed out that the experimental data can
be equally well fitted with both these expressions for $Q_s$. This is also seen
in fig.~\ref{fig:DLA}, which shows the cross
section $\sigma^{\gamma^*p}\sim g/Q^2$ \emph{vs.} $\tau$ with $g$ from eq.~(\ref{eq:DLLA})
and $\tau$ defined from eq.~(\ref{eq:R0}) but with $\lambda = 0.7$. 
\FIGURE[t]{
  \includegraphics[angle=270,  scale=0.8]{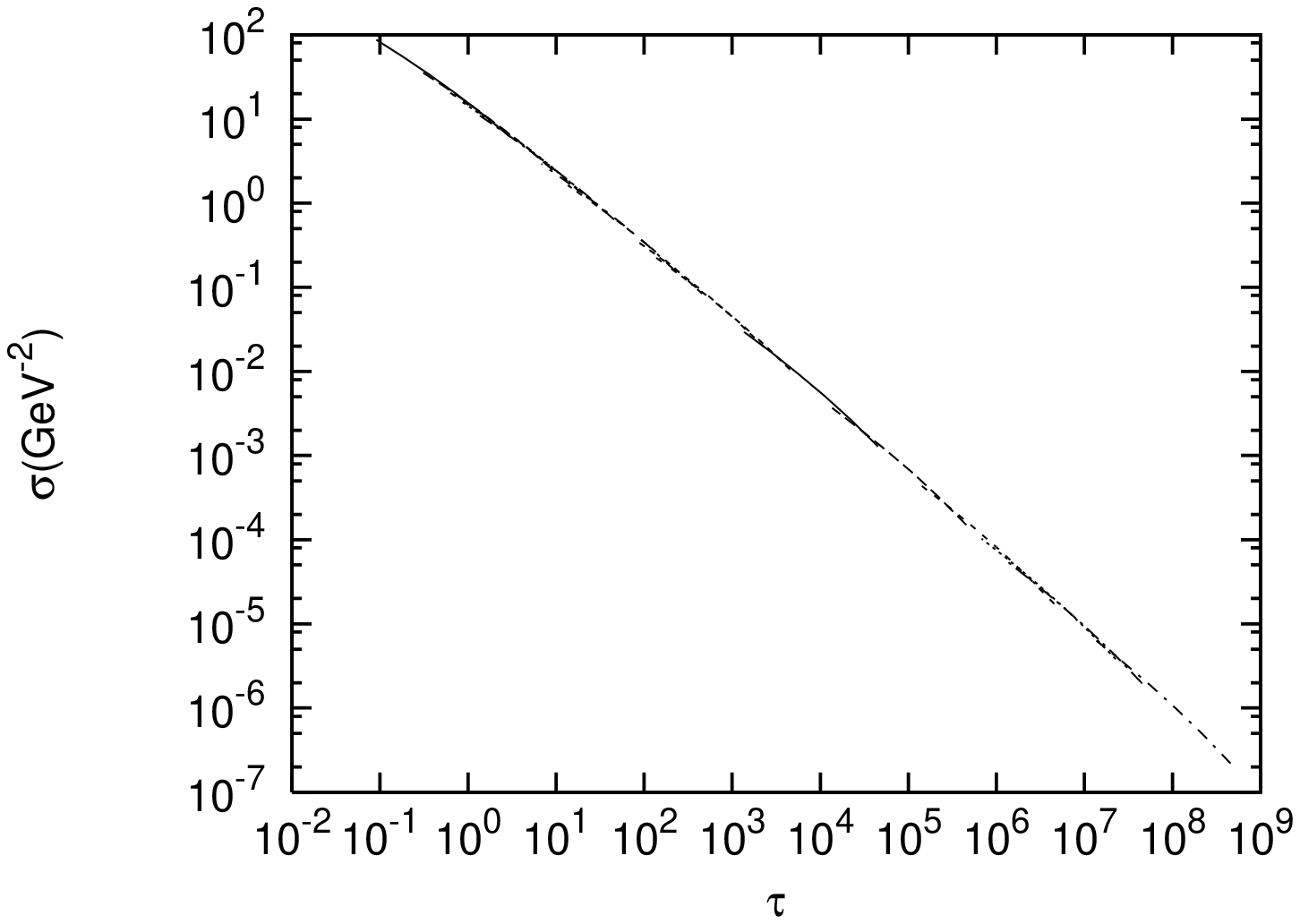}
  \caption{\label{fig:DLA} The cross section $\sigma_{\gamma^*p}\sim g/Q^2$
    with the gluon density $g$ given by equation \eqref{eq:DLLA} for different 
    $Q^2$, varying from $Q^2=100$GeV$^2$ to $Q^2=1.5\cdot 10^{10}$GeV$^2$, and plotted 
    as a function of $Q^2/Q_s^2$, with $Q_s$ parametrized according to 
    \eqref{eq:R0}, but with $\lambda$ equal to 0.7. } }

For limited $Q^2$ we are instead in the BFKL regime dominated by 
$k_\perp$-non-ordered chains, where the gluon density is growing 
as a power $1/x^\lambda$ for small $x$, 
with $\lambda$ of the order 0.3. Multiplying by $1/Q^2$ 
therefore gives directly the scaling cross section
\begin{equation}
\sigma^{\gamma^* p} \sim \frac{Q_s^2}{Q^2} = \tau^{-1}
\label{eq:bfkl}
\end{equation}
with $Q_s^2 \propto 1/x^{0.3}$.

We now also want to argue that, as described in ref.~\cite{Gustafson:2001iz},
the line $Q^2 = Q_{\mathrm{limit}}^2$  
corresponding to the separation between ordered (DGLAP-like) 
chains and unordered (BFKL-like) chains, is also close to $Q_{s,GW}^2$. 
In the double log approximation the (non-integrated) gluon distribution $G$
is given by the $k_\perp$-ordered expression
\begin{equation}
G(x,k_\perp^2) \sim \sum_n \prod_i^n \left\{ \int \frac{4\alpha_s}{3\pi} 
\frac{d k_{\perp,i}^2}{k_{\perp,i}^2}
\theta(k_{\perp,i} - k_{\perp,i-1})
\frac{d x_i}{x_i}\theta(x_{i-1} - x_i)
\right\} 
\, \delta(x- x_n) \delta(k_\perp^2 - k_{\perp,n}^2).
\label{eq:smallxchain}
\end{equation}
With the notation
%\begin{equation}
$l_i \equiv \ln (1/x_i)$ and
%l \equiv \ln (1/x);\,\,\,
$\kappa_i \equiv \ln (k_{\perp,i}^2/\Lambda^2)$
%\kappa \equiv \ln (q_{\perp}^2/\Lambda^2);\,\,\, 
%\label{defkappal}
%\end{equation}
(and $\kappa = \ln (k_{\perp}^2/\Lambda^2),\, l = \ln (1/x)$)
we get for a fixed coupling the result
\begin{eqnarray}
G &\sim& \sum_n \left\{ \prod_i^n \int^{\kappa} \bar{\alpha} d\kappa_i 
\theta(\kappa_i - \kappa_{i-1})   
\cdot \prod_i^n \int^{\ln 1/x} dl_i 
\theta(l_i - l_{i-1}) \right\} \nonumber \\
&=&\sum_n \bar{\alpha}^n \cdot \frac{\kappa^n}{n!}\cdot\frac{l^n}{n!} \,
= \,I_0(2\sqrt{\bar{\alpha} \ln Q^2 \ln 1/x})
\sim \exp\left(2\sqrt{\bar{\alpha} \ln Q^2 \ln 1/x}\right).
\label{eq:DLL}
\end{eqnarray}
For a running $\alpha_s$ we have instead of 
$d k_{\perp,i}^2 /k_{\perp,i}^2 = d \kappa_i$ a factor 
$d\kappa_i/\kappa_i = d \ln \kappa_i$, which then gives the result
in eq.~(\ref{eq:DLLA}). 

In the BFKL region with small $x$ but not so large $Q^2$, the 
$k_\perp$-ordered phase space in eq.~(\ref{eq:smallxchain}) 
becomes small, and chains which are not
ordered in $k_\perp$ give important contributions. The BFKL evolution
can be formulated in different ways. Expressed in terms of the
primary \cite{Andersson:1995ju} or backbone \cite{Salam:1999ft} gluons
a step downwards in $k_\perp$ is suppressed
by a factor $k_{\perp i}^2/k_{\perp i-1}^2$ \cite{Andersson:1995ju,Gustafson:2003nu}.
We note that this implies that a maximum $k_\perp$-value in the chain
will contain the factor $d k_{\perp \mathrm{max}}^2/k_{\perp \mathrm{max}}^4$
which can be interpreted as a hard parton-parton subcollision
with the expected cross section proportional to $d \hat{t}/\hat{t}^2$.
Expressed in the logarithmic variable $\kappa$, a step
down is consequently suppressed by a factor 
$\mathrm{exp}(\kappa_i-\kappa_{i-1})=\mathrm{exp}(-\delta\kappa)$. 
This 
implies that the effective range allowed for downward steps corresponds 
to approximately one unit in $\kappa$. Consequently we find that the 
phase space limits $\kappa_i \gtaet \kappa_{i-1}$ in eq.~(\ref{eq:DLL})
is replaced by \cite{Gustafson:2001iz,Gustafson:2003nu}
\begin{equation}
\kappa_i \gtaet \kappa_{i-1} - 1.
\label{eq:bfklfasrum}
\end{equation}

For a fixed $\alpha_s$ the transverse momentum integrals giving
$\kappa^n/n!$ 
in eq.~(\ref{eq:DLL})
%\begin{equation}
%\int_0^\kappa \prod_i^n d\kappa_i \theta(\kappa_i - \kappa_{i-1}) = 
%\frac{\kappa^n}{n!}
%\label{fixalfa}
%\end{equation}
%
will be replaced by 
\begin{equation}
\int_0^\kappa \prod_i^n d\kappa_i \theta(\kappa_i - \kappa_{i-1} - 1) \approx 
\frac{(\kappa+n)^n}{n!} 
\label{fixalfasmallx}
\end{equation}
When $\kappa$ is very large we recover the DLL result in eq.~(\ref{eq:DLL}), 
but for smaller values of $\kappa$ we find instead using Stirling's formula
\begin{equation}
\kappa \,\,\mathrm{small}\,\,\Rightarrow \,\,\frac{(\kappa+n)^n}{n!} 
\sim \frac{n^n}{n!} \sim e^n
\label{eq:kappalitet}
\end{equation}
which implies
\begin{equation}
G \sim \sum_n \frac{[ \bar{\alpha}\, e \ln(1/x)]^n}{n!} = 
e^{e \,\bar{\alpha} \ln(1/x)} = \frac{1}{x^\lambda}
\label{xlambda}
\end{equation} 
with 
\begin{equation}
\lambda = e \,\bar{\alpha} \approx 2.72 \,\bar{\alpha}.
\label{approxlambda}
\end{equation}
In this range the chain corresponds to a random walk in $\ln k_\perp^2$. 
The result should be compared with the result from the leading order 
BFKL equation, which gives
\begin{equation}
\lambda = 4\ln2 \,\bar{\alpha} \approx 2.77 \,\bar{\alpha}.
\label{bfkllambda}
\end{equation}
We see that this simple picture describes the essential 
features of BFKL evolution.

We note here in particular that the boundary between the domains
dominated by $k_\perp$-ordered chains (eq.~(\ref{eq:DLL})) and 
$k_\perp$-non-ordered chains (eq.~(\ref{eq:kappalitet})) is determined
by the relation
\begin{equation}
\ln k_{\perp \mathrm{limit}}^2 = e \,\bar{\alpha} \ln(1/x).
\end{equation}
We see that if we replace the LL $\lambda$-value by a lower value 
$\approx 0.3$, as indicated by the experimental data, then
$k_{\perp \mathrm{limit}}^2$ is similar to $Q_{s,GW}^2(x)$
from the early fit by Golec-Biernat and W\"usthoff 
\cite{Golec-Biernat:1998js,Golec-Biernat:1999qd}. This shows 
that the line $Q^2 = Q_{s,GW}^2(x)$ (corresponding to $\tau = 1$) 
is actually close to $Q_{\mathrm{limit}}^2$, which represents 
the separation between $k_\perp$-ordered (DGLAP-like) 
chains and non-ordered (BFKL-like) chains.

\subsection{The dipole cascade}
\label{sec:linearcascade}

The qualitative features in the LLA in sec.~\ref{sec:scaleLL}
are modified by energy conservation and other non-leading effects.
We will here study how the scaling behaviour can be seen in the 
dipole cascade model presented in sec.~\ref{sec:cascademodel},
trying now to isolate the most important features of
the full MC simulation.

Let us look at a dipole with size $r$, which after a rapidity interval 
$\Delta y$ splits in two dipoles with sizes $r_>$ and $r_<$,
with $r_>$ larger than $ r_<$.
Fig.~\ref{fig:adistrb} shows the MC results for the average values 
of the ratios $r_>/r$ and $r_</r$ for different values for the ``time'', 
$\Delta y$, it takes from the formation of the dipole till it splits.
The probability distribution $dP/d\Delta y$ is also shown.
We see here that a dipole typically splits in two
dipoles, where one has the same size as the parent while
the other has just half its size. This result is independent of the 
step $\Delta y$ and of the photon virtuality $Q^2$. This also shows
that it is independent of the size of the parent dipole.
We note that these results depend crucially on the
energy-momentum conservation in the evolution. Without this constraint 
the typical splitting would occur within a very small rapidity step
giving one virtual dipole with a size close to the necessary cutoff.
When the cutoff goes to zero both $\Delta y$ and $r_</r$ would also go to zero. 
Essential for the simple result 
in fig.~\ref{fig:adistrb} is the conservation of both $p_+$ and $p_-$. This is
seen in fig.~\ref{fig:adistrbnopmin}, which shows the corresponding result 
obtained with only $p_+$-conservation.

Let us then assume that each dipole $r$ splits after a typical $\Delta y$, 
which from the distribution $dP/d\Delta y$ is found to be around 1.8 units,
into two dipoles with sizes $r$ and $r/a$, with the parameter $a$ of the 
order of 2. If $Y$ denotes the total rapidity range the number of steps
will be $N=Y/\Delta y$ and  
the number of dipoles $2^N=2^{(Y/\Delta y)}$. 
Starting from an initial dipole $r_0$ the number of dipoles
having size $r_0/a^n$ will be equal to $\binom{N}{n}$, with $n = 0, 1, \dots, N$.  

\FIGURE[t]{
  \includegraphics[angle=270, scale=0.5]{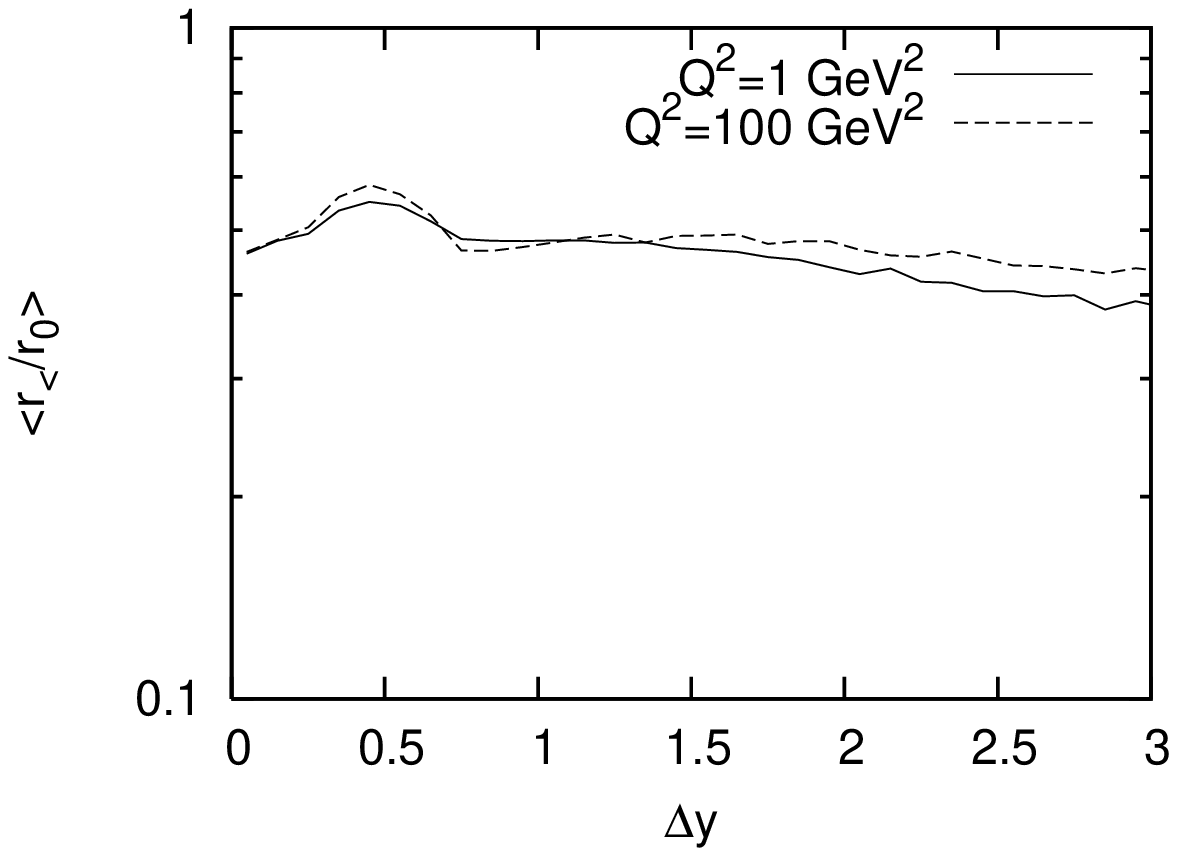}
  \includegraphics[angle=270, scale=0.5]{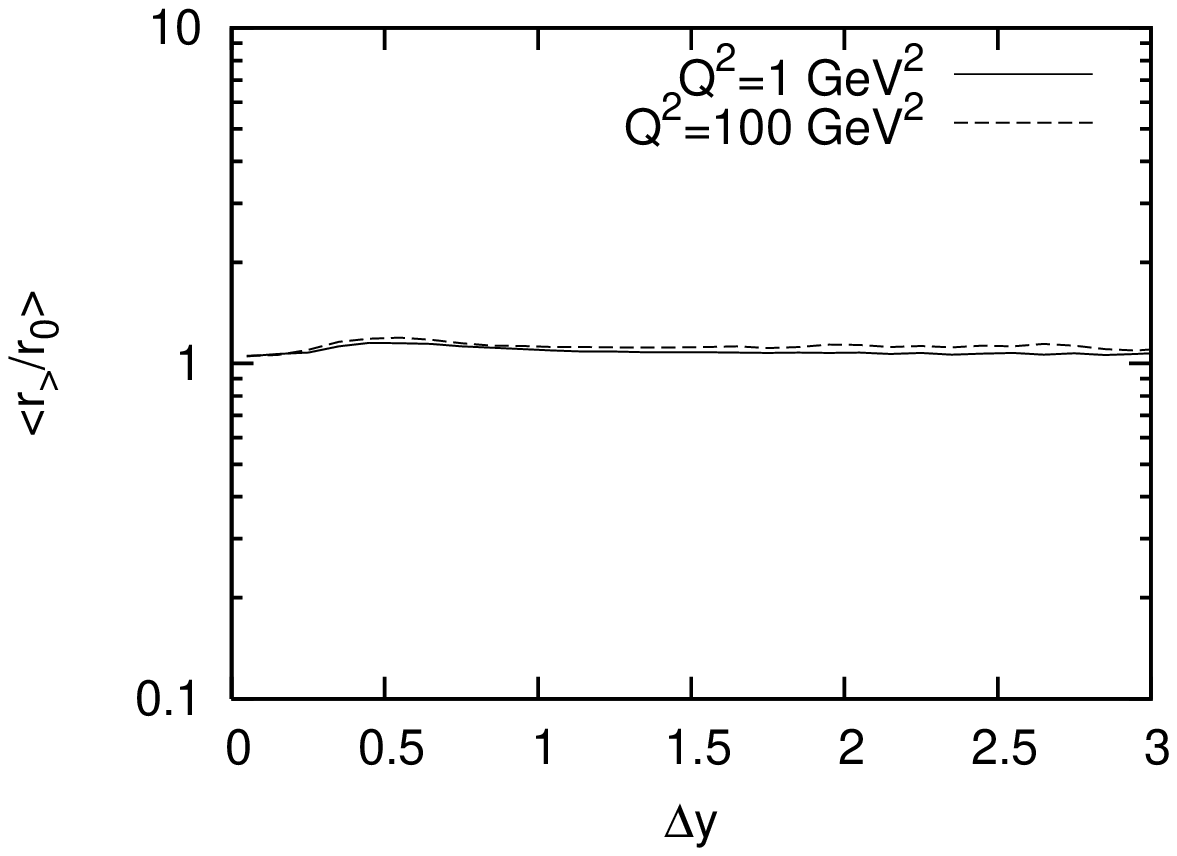}
  \includegraphics[angle=270, scale=0.5]{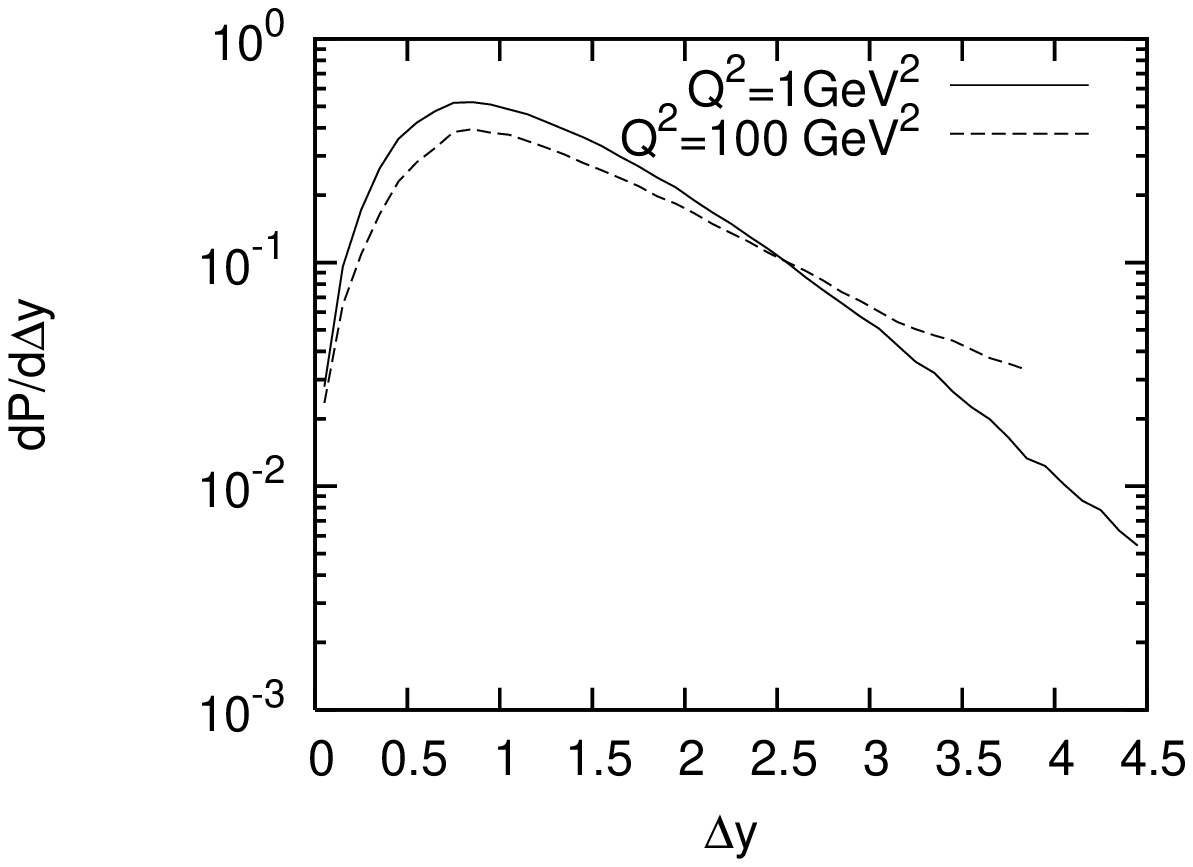}
  \caption{\label{fig:adistrb} The top figures show how the average values of 
    the ratios
    $r_</r_0$ (left) and $r_>/r_0$ (right) for splitting events vary with 
    the rapidity separation $\Delta y$. 
    It is seen that both these
    ratios are approximately constant independent of $\Delta y$, and that $r_>/r_0$ 
    is $\approx 1$. The bottom figure presents the $\Delta y$-distribution
    $dP/d\Delta y$. This distribution has a peak around $\Delta y=1$ and falls 
    off exponentially for larger $\Delta y$-values. } }

To study the scaling behaviour of the $\gamma^*p$ cross section 
within this approximation, we can treat the photon as a
dipole with size $r_\gamma=1/Q$. In the linear region
the initial proton can also be treated as a single dipole with size
$r_p=1/\Lambda$. 
The cross section for the scattering of two dipoles, $r_1$ and 
$r_2$, is given by \cite{Mueller:1994gb}
\begin{eqnarray}
\sigma_{dd}(r_1,r_2)=2\pi \alpha_s^2 r_{min}^2\{1+\mathrm{ln}(\frac{r_{max}}{r_{min}})\}
\end{eqnarray}
where $r_{max}$ ($r_{min}$) is the largest (smallest) of the two colliding 
dipoles $r_1$ and $r_2$. We study the collision in a frame 
where only the proton is evolved, and the photon therefore treated as a single
dipole. 

\FIGURE[t]{
  \includegraphics[angle=270, scale=0.5]{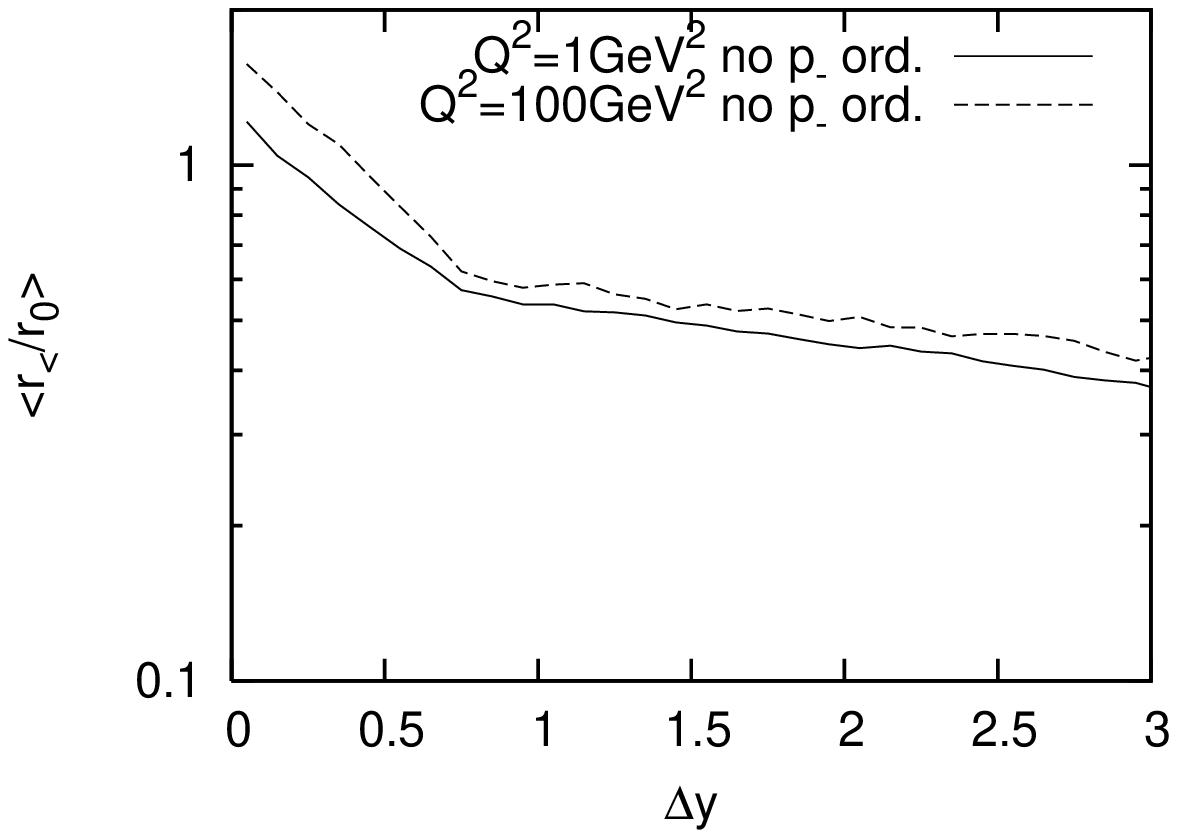}
  \includegraphics[angle=270, scale=0.5]{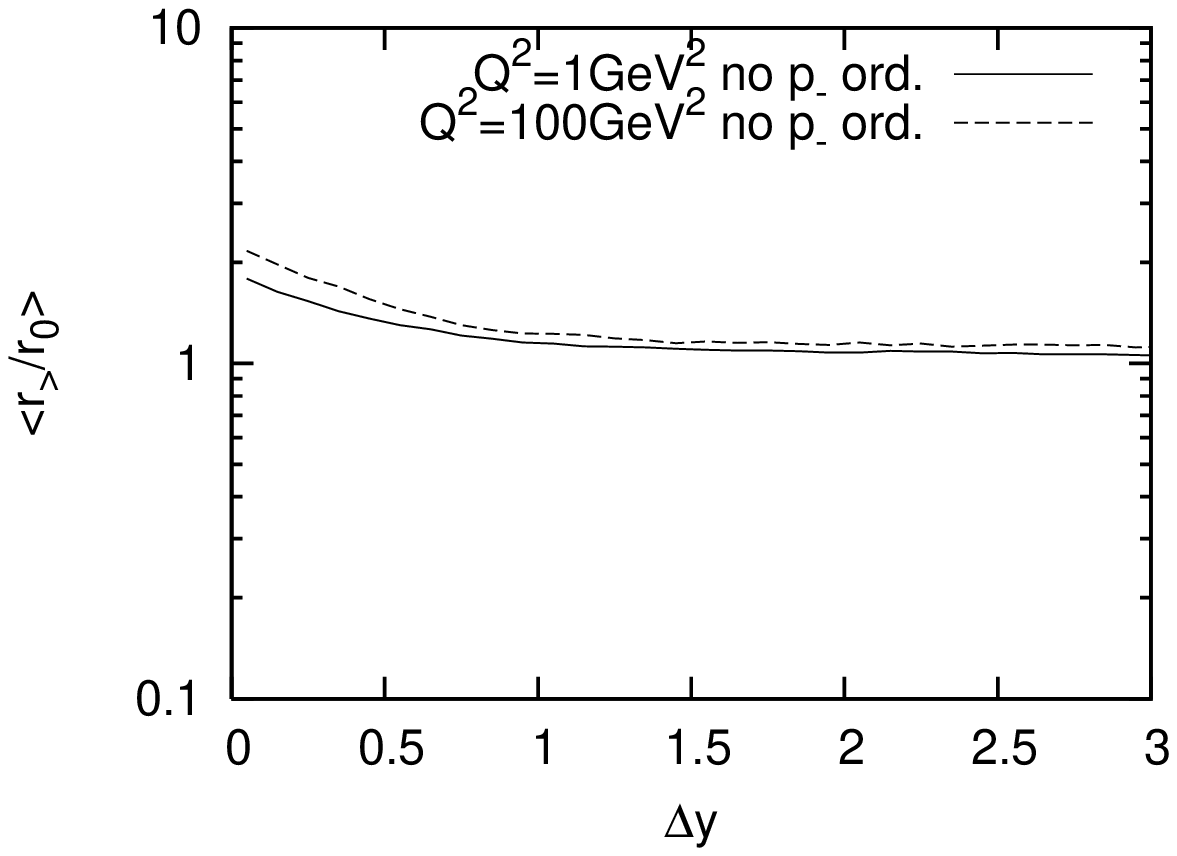}
  \caption{\label{fig:adistrbnopmin} The averaged ratios
    $\langle r_</r_0\rangle$ and $\langle r_>/r_0\rangle$ 
    as functions of $\Delta y$ in a simulation where $p_-$ is not conserved in the 
    evolution. We see that $p_+$-conservation is not enough to get the 
    almost constant ratios seen in figure \ref{fig:adistrb}. } }

When $Q$ is larger than $a^N \!\cdot\!\Lambda$ the photon dipole is smaller
than all dipoles in the proton cascade. The resulting 
cross section is then given by
\begin{eqnarray}
\sigma^{\gamma^*p}(Q^2,Y)& =& 2\pi \alpha_s^2 \sum_{n=0}^N \binom{N}{n} 
Q^{-2}\left\{1+\mathrm{ln}(\frac{Q}{a^n\Lambda})\right\} \nonumber \\
&=& \pi \alpha_s^2 Q^{-2}2^{N}
\left\{2-N\mathrm{ln}a+\mathrm{ln}\frac{Q^2}{\Lambda^2}\right\} 
\propto \frac{Q_{sc}^2}{Q^2} \left\{1+\frac{1}{2}\ln(\frac{Q^2}{Q_{sc}^2})\right\},
\label{eq:toysigma}
\end{eqnarray}
where in the last expression we have used $a=2$, $N=Y/\Delta y$ and
$Q_{sc}^2 = \Lambda^2x^{-\ln2/\Delta y}\approx \Lambda^2x^{-0.38}$. The result is obviously  
scaling, and the exponent 0.38 is not far from the experimental fit around 0.3.
We note also that in this case the
dominating contribution comes from dipole chains where the dipoles are ordered
in size and where the last dipole in the cascade is larger than
the photon dipole. This just corresponds to the dominance of $k_\perp$-ordered
DGLAP-type ladders.  

For smaller $Q$-values the curly bracket in eq.~\eqref{eq:toysigma} 
overestimate the contribution from small proton dipoles represented by
$n$-values for which $a^n \!\cdot\!\Lambda > Q$. The largest terms in the 
sum are obtained when the binomial factor has its maximum, \emph{i.e.} for
$n\approx N/2$. Therefore the result in  eq.~\eqref{eq:toysigma} is
a good approximation as long as $Q >a^{N/2}\!\cdot\!\Lambda$. 
Indeed, using Stirling's formula 
we can write $\binom{N}{N/2}\approx 2^N$, and in the saddlepoint approximation
we again arrive at the result given in eq.~\eqref{eq:toysigma}. 
We note that the constraint $Q >a^{N/2}\!\cdot\!\Lambda$ corresponds 
just to $Q^2>Q_{sc}^2$, which thus is the limit for the dominance
of $k_\perp$-ordered, DGLAP-type evolution chains.

%\begin{eqnarray}
%\sigma^{\gamma^*p}(Q^2,Y) &\sim& \frac{2\pi \alpha_s^2 }{Q^2}
%2^N \left\{1+\frac{1}{2}\ln(\frac{Q^2}{a^{N}\Lambda^2})\right\}=
%\nonumber\\
%&=& 
% \frac{2\pi \alpha_s^2 }{Q^2}\exp\left(\frac{\ln 2}{\Delta y}Y\right)
%\left\{1+\frac{1}{2}\ln(\frac{Q^2}{\Lambda^2 /x^{\ln a/\Delta y}})\right\}
%\nonumber\\
%&\propto& \frac{Q_{sc}^2}{Q^2} \left\{1+\frac{1}{2}\ln(\frac{Q^2}{Q_{sc}^2})\right\}
%\label{eq:toysigma2}
%\end{eqnarray}
%which is identical to eq.~\eqref{eq:toysigma}.

Below this region, \emph{i.e.} for $Q^2 < Q_{sc}^2$, it is not a 
good approximation to 
neglect the contributions from scatterings where the proton dipoles 
are smaller than the photon dipole. The full expression, including all 
contributions, can be written as 
\begin{eqnarray}
\sigma^{\gamma^*p}(Q^2,Y) \propto \sum_{n=0}^m \binom{N}{n} 
Q^{-2}\left\{1+\mathrm{ln}(\frac{Q}{a^n\Lambda})\right\} + 
\sum_{n=m+1}^N \binom{N}{n}a^{-2n}\Lambda^{-2}\left\{1+\mathrm{ln}(
\frac{a^n\Lambda}{Q})\right\} \nonumber \\
\label{eq:toysigma2}
\end{eqnarray}
where $m\equiv\frac{\mathrm{ln}Q/\Lambda}{\mathrm{ln}a}$.
This expression is more complicated, but we show in figure \ref{fig:appdixfig2}
the result of a numerical evaluation expressed with a scale 
$Q_{sc}^2 \propto x^{-0.4}$. We see that scaling is indeed
satisfied for a large range of values for $Q^2$, also in the kinematic region
dominated by chains which are not well ordered in dipole size
or in transverse momentum.

\FIGURE[t]{
  \includegraphics[angle=270,  scale=0.8]{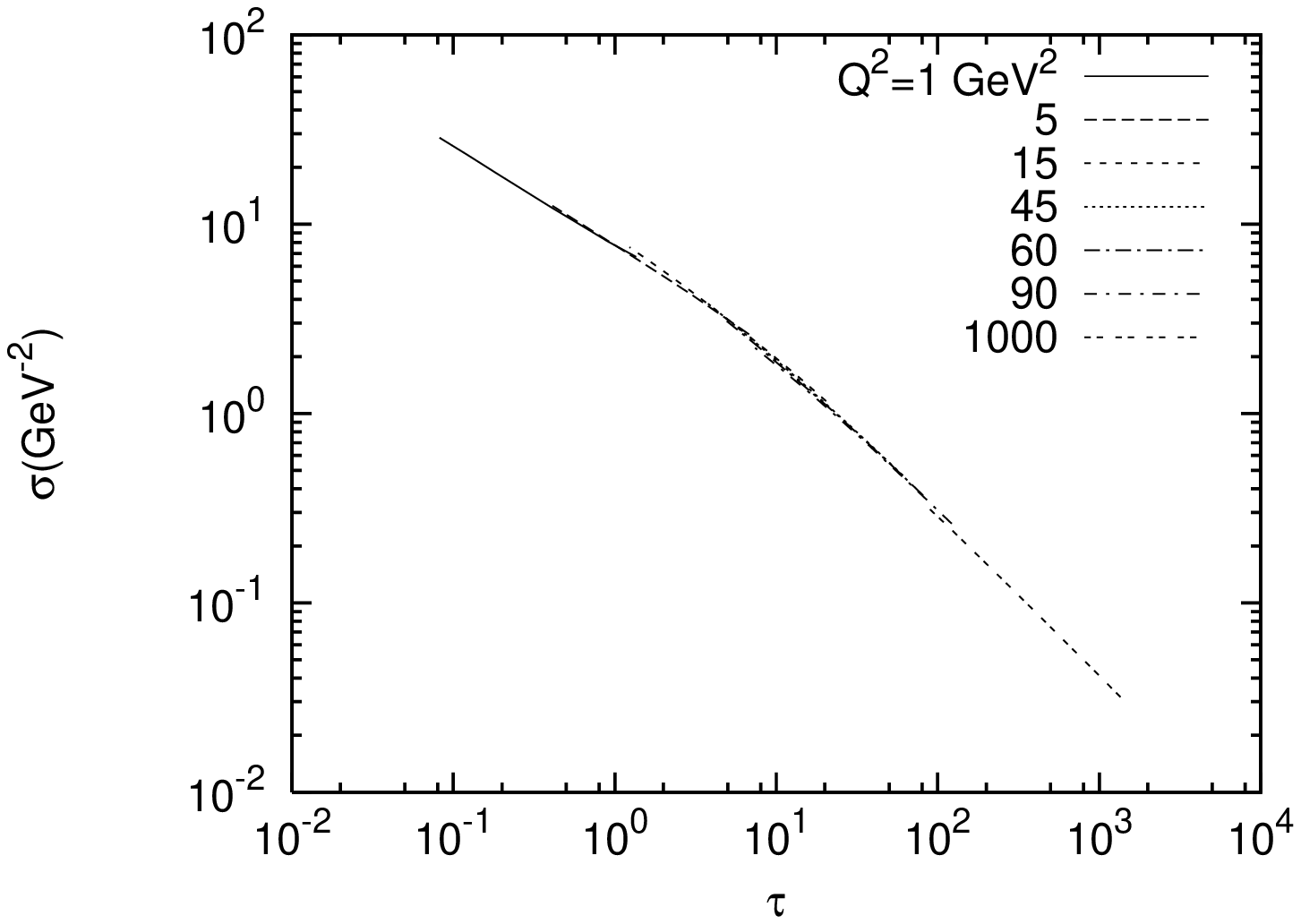}
  \caption{\label{fig:appdixfig2} The toy model cross section in 
    \eqref{eq:toysigma2} plotted as a function of the scaling variable
    with $Q_{sc}^2 = \Lambda^2\cdot x^{-0.4}$, with $\Lambda^2 = 0.1$GeV$^2$, 
    and for different $Q^2$. The normalization here is not of interest and all 
    prefactors has been simply put to 1. We indeed see that the result shows 
    scaling, for low and high $Q^2$ alike. }
}

We note that our toy model has important similarities with the 
empirical model presented in ref.~\cite{Munier:2002gf}. 
In this paper it is demonstrated that geometric scaling follows from
an assumption that the dipole
cascade in the proton is dominated by dipoles with size $Q_s^{-1}$ (while
the virtual photon can be represented by a dipole with size $Q^{-1}$).
Our toy model approximation of the full MC cascade has just this feature; 
the dipole multiplicity is given
by the binomial coefficient $\binom{N}{n}$, with a maximum for
$n=N/2$ corresponding to $r=1/Q_{sc}$. The result in eq.~\eqref{eq:toysigma},
which is obtained for $Q>Q_{sc}$ using the saddlepoint approximation,
is thus identical to the corresponding result in \cite{Munier:2002gf}.
Ref.~\cite{Munier:2002gf} also points out that if
events with $Q<Q_s$ can be described by an analogous cascade evolution
of the initial photon dipole, then the expression 
$Q/Q_{s}\sigma^{\gamma^*p}(Q/Q_{s}(x))$ is 
symmetric under the exchange $Q \leftrightarrow Q_{s}(x)$. This
symmetry was initially observed in the HERA data by ref.~ \cite{Stasto:2000er}.
For these smaller $Q$-values ($Q<Q_{sc}$) the dominant contributions
from $n\approx N/2$ are contained in the second term in 
eq.~\eqref{eq:toysigma2}. Using again the saddlepoint 
approximation we also obtain the symmetric result
\begin{eqnarray}
\sigma^{\gamma^*p}(Q^2,Y) \approx 2\pi\alpha_s^2 2^Na^{-N}\Lambda^{-2}\left\{1 + \mathrm{ln}
(\frac{a^{N/2}\Lambda}{Q})\right\} \approx 2\pi\alpha_s^2 \Lambda^{-2}
\left\{1+\mathrm{ln}(\frac{Q_{sc}}{Q})\right\}.
\label{eq:toysigma3}
\end{eqnarray}
    
In conclusion we find that $Q^2 = Q_{sc}^2\propto x^{-\lambda}$, 
with $\lambda$ approximately equal to the BFKL exponent, specifies 
the limit between dominance of
$k_\perp$-ordered and $k_\perp$-non-ordered chains, and that a simple
toy model having this property gives the qualitative features of the 
scaling dynamics for both large and small $\tau$. We also see that 
the toy model approximation to the full MC simulation of the dipole cascade
model have important similarities with the empirical model in 
ref.~\cite{Munier:2002gf}, and gives a scaling result with a
scaling exponent not far from what is observed at HERA.

\section{Scaling features in the charm contribution}
\label{sec:charm}

It is well known from 
HERA data that charm quarks contribute a significant part to the 
total cross section.
As discussed in sec.~\ref{sec:saturation}, the large charm quark mass 
modifies the scaling properties, and it is seen that  
the HERA charm data do indeed not scale as a function of $\tau=Q^2/Q_{s,GW}^2(x)$.
As pointed out in ref.~\cite{Goncalves:2006ch} they do, however, scale quite
well as a function of the modified scaling variable 
$\tau_c=(Q^2+4m_c^2)/Q_s^2$

From the photon wave function in \eqref{eq:psigamma} we see that the term
proportional to the Bessel function $K_1^2$ in $|\psi_T|^2$ only contains
$Q^2$ and $m_f$ in the combination $z(1-z)Q^2 + m_f^2$. If $z$-values around 
1/2 dominate, we would expect that the charm
contribution is gradually switched off when $Q^2$ is of the order $4 m_c^2$
or smaller, which necessarily leads to a breaking of geometric scaling.  
The sum of the terms proportional to $K_0^2$ in $|\psi_T|^2$ and $|\psi_L|^2$
is proportional to $[4z(1-z)]z(1-z)Q^2+m_f^2$, and for $z$ close to 1/2
the square bracket equals 1, and we get the same factor $z(1-z)Q^2 + m_f^2$
as before. From these features we may expect
that the charm contribution scales approximately with a 
modified scaling parameter where, for example, we replace $Q^2$ by 
$Q^2+n\!\cdot\! m_c^2$ in the definition of $\tau$, for some number $n$
which should be close to or a little larger than 4. 

In fig.~\ref{fig:charmscaling} 
we show data for the charm cross section from \textsc{Zeus} 
\cite{Chekanov:2003rb} and our MC obtained for $m_c=1.4 \,\mathrm{GeV}$.
In this figure we have used the scaling variable
\begin{equation}
\tau_c=(Q^2+6 m_c^2)/Q_{s,GW}^2(\bar{x}) \,\,\,\, \mathrm{with} \,\,\,\, \bar{x}=(Q^2+6 m_c^2)/W^2,
\label{eq:charmscale}
\end{equation}
obtained with $n=6$ in the definition of $\tau_c$, which we find gives
somewhat better scaling properties. This
implies that $\tau_c=(1+6 m_c^2/Q^2)^{1+\lambda}\!\cdot\!\tau_{GW}$.
We here first note that the MC agrees quite well with the data, being 
only a little high for the largest $\tau$-values.
Secondly we confirm the observation in ref.~\cite{Goncalves:2006ch}
that the charm cross section does scale well
as a function of such a modified scaling variable. 

The value $n=6$ chosen for the scaling variable in eq.~\eqref{eq:charmscale}
is not crucial for the scaling 
behaviour, which works rather well also for $n=4$. 
We note, however, that replacing $x$ by $\bar{x}$ in $Q_s$ only contributes
a minor part to the difference between $\tau_c$ and $\tau_{GW}$, represented by
the factor $(Q^2+6 m_c^2)^{\lambda}$. More important is the replacement
$Q^2\rightarrow Q^2+6 m_c^2$ in eq.~\eqref{eq:charmscale}. This is
illustrated in fig. \ref{fig:charmscaling2}, which shows that scaling
does not hold if we replace $\tau$ by $Q^2/Q_{s,GW}^2(\bar{x})$. 

\FIGURE[t]{
  \includegraphics[angle=270, scale=0.8]{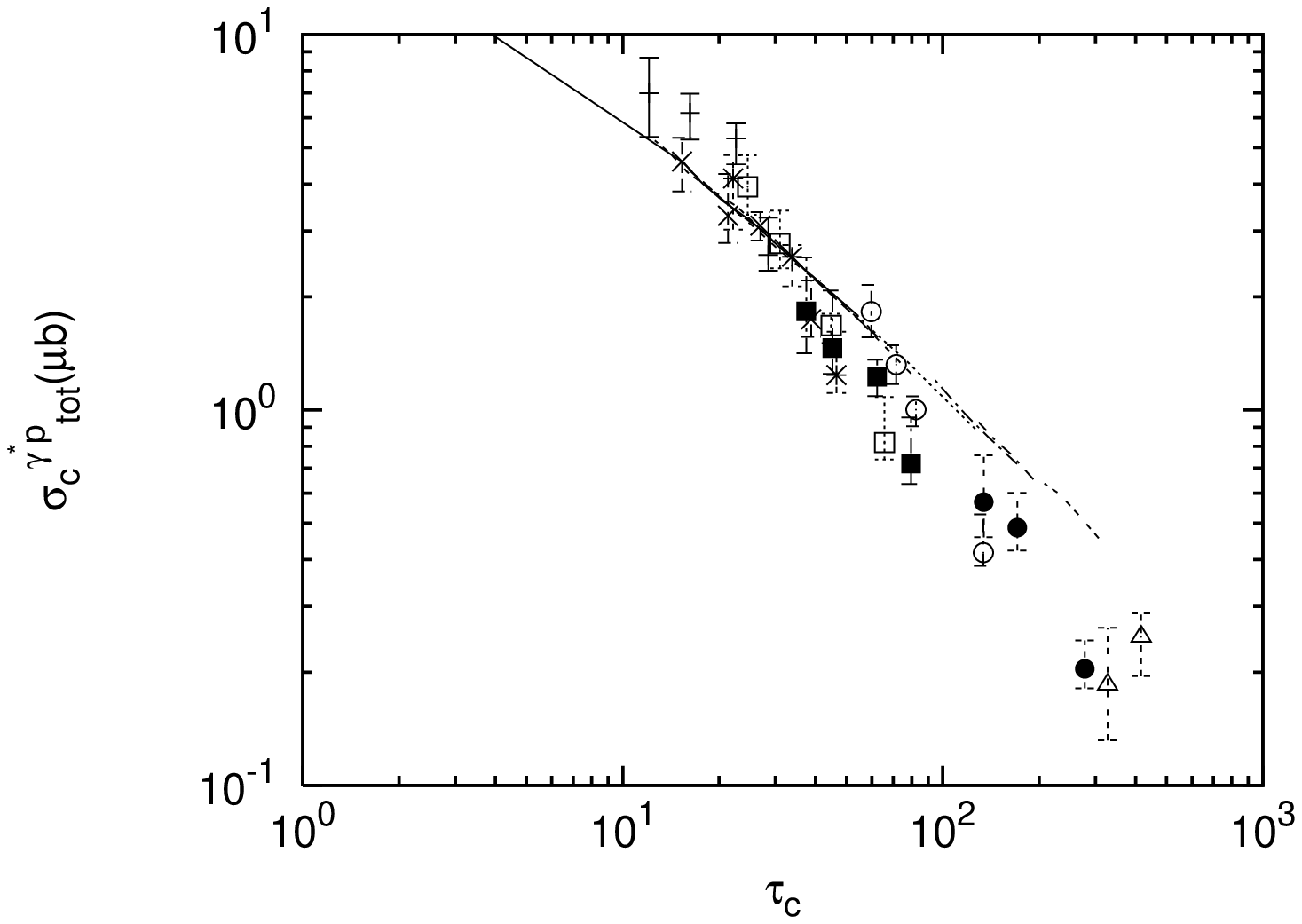}
  \caption {\label{fig:charmscaling} The total charm cross section, $\sigma_T^c+
    \sigma_L^2$ plotted as a function of $\tau_c$ defined in \eqref{eq:charmscale} 
    with $\lambda =0.35$ and $m_c =1.4$ GeV.
    Results are shown for $Q^2$ between 2 and $130$ GeV$^2$. 
    We see that the result scales fairly well with this scaling parameter. 
    Data points are taken from ref \cite{Chekanov:2003rb}. } }

We want here to point out that the effect of the charm mass is reduced
very slowly for larger $Q^2$.
Integrating over the dipole size, $\mathbf{r}$, we obtain from a dimensional analysis  
\begin{eqnarray}
\int d^2\pmb{r} K_i^2(\epsilon r) \sigma(z,\pmb{r}) \propto \frac{1}{\epsilon^4}\,, 
\qquad \epsilon = \sqrt{z(1-z)Q^2+m_f^2}.
\end{eqnarray}
When $4 Q^2$ is small compared to $m_f^2$ we have $\epsilon \approx m_f$, 
%$Q^2 \lesssim 4m_c^2$, which implies that $z(1-z)Q^2 \leqslant 0.25Q^2 \lesssim m_c^2$,
and the charm cross section ought to scale as 
\begin{eqnarray}
\sigma_T \propto \frac{1}{m_c^2}, \qquad \sigma_L \propto \frac{Q^2}{m_c^4}.
\end{eqnarray}
We see here that the longitudinal contribution depends on two separate scales 
$Q^2$ and $m_c^2$. 
In the other limit, when $z(1-z)Q^2 > m_c^2$ 
we can neglect $m_c$. However, since $z(1-z)$ can take on
arbitrarily small values for fixed $Q^2$, we the effect of the
charm mass can be important also for high $Q^2$.
In the MC results presented in fig.~\ref{fig:charmscaling} the reduction
due to the charm mass is about 30\% for $Q^2 = 90\, \mathrm{GeV}^2$
and 55\% for $Q^2 = 15\, \mathrm{GeV}^2$. This slow decrease of the mass
effect is also seen in fig.~\ref{fig:mcscaling2}.
(We should here also remark that 
the transverse wave function $\psi_T$ is not normalizable due to the 
singularity $K_1^2(r) \sim 1/r^2$ at small $r$. However, at small $r$ the 
dipole-proton cross section behaves like $r^2$ and thus the result is still finite.)

\FIGURE[t]{
  \includegraphics[angle=270, scale=0.8]{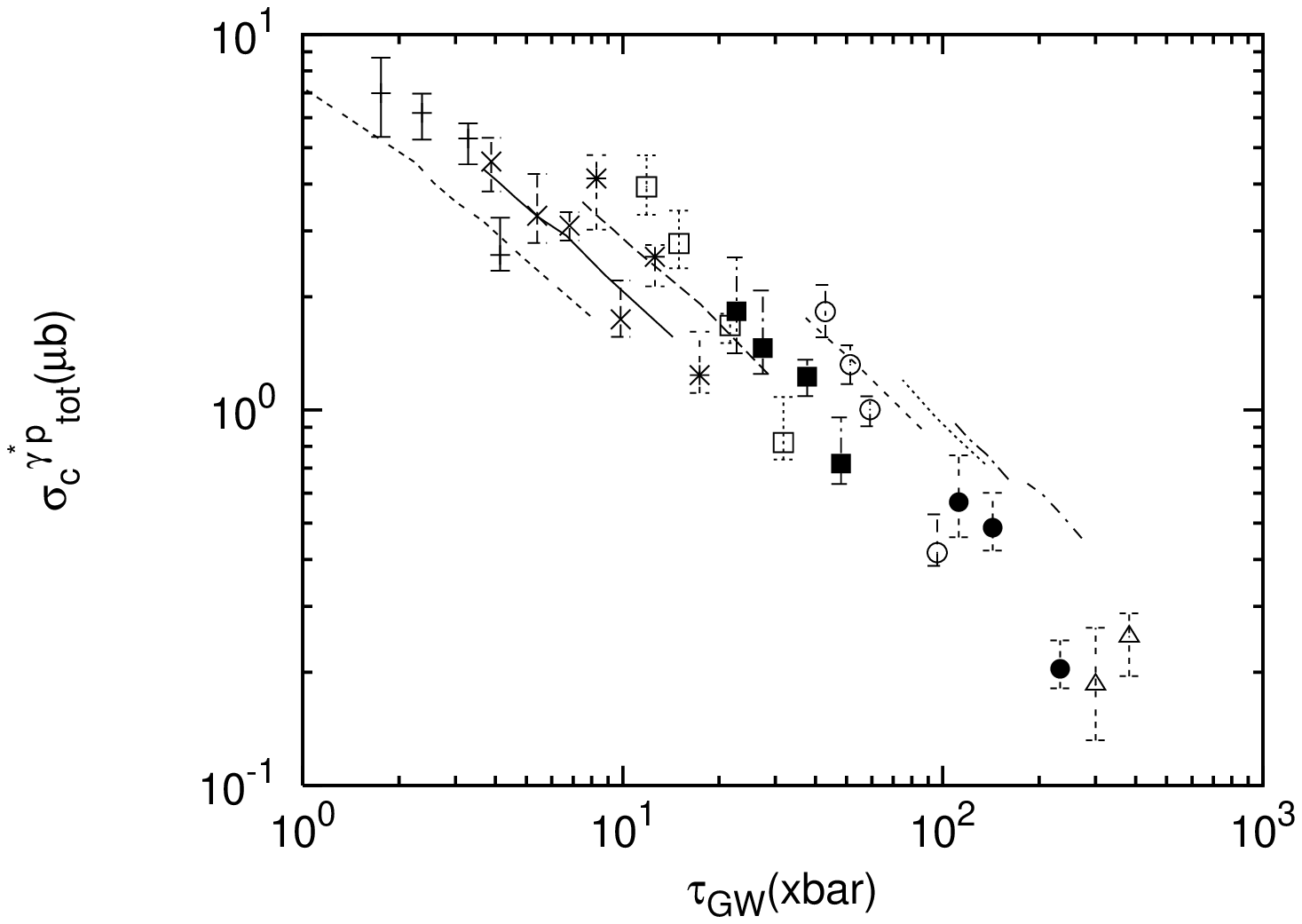}
  \caption {\label{fig:charmscaling2} Here we plot the charm cross section 
    as a function of $Q^2/Q_s^2(\bar{x})$ with the same parameters 
    as in fig. \ref{fig:charmscaling}.
    In this case we see that the result does not scale, which shows the 
    importance of replacing $Q^2$ with $Q^2+n\cdot m_c^2$, $n \sim 4$, 
    as argued in the text.  } } 

\section{Interaction at smaller $Q^2$}
\label{sec:smallQ}

Due to the limited energy in the HERA accelerator
the experimental data in the region $\tau<0.5$ are all obtained
for rather small virtualities $Q^2$, where non-perturbative effects must be expected. 

\subsection{Can the perturbative dipole formalism be used for $Q^2$ below 1 GeV$^2$?}

For small $Q^2$ the wave functions in eq.~(\ref{eq:psigamma}) extend to
very large transverse separations $r$, and such large dipoles must be
suppressed by confinement effects. The interaction of these photons is 
usually described in terms of two separate components, a vector dominance 
contribution and a direct coupling to the quarks. We will here test if it is 
possible to represent the interaction of these photons with a finite
mass for the light quarks, which effectively suppresses the contribution from the
very large dipoles. Because the contribution from the strange quark is 
suppressed by its smaller electric charge, and therefore relatively small,
it is not possible to study in any detail the effect of the strange 
quark mass. We therefore use a single 
quark mass, $m_l$, for all the light quarks, $u$, $d$, and $s$.
The results in fig.~\ref{fig:smallQ} for $Q^2$ in the range 0.3-3.5 
$\mathrm{GeV}^2$ are obtained for $m_l=60 \mathrm{MeV}$ and
$m_c=1.4 \mathrm{GeV}$, and we here show the model results for
the same combinations of $x$- and $Q^2$-values as in the experimental data.
We see that the experimental data are very
well reproduced by the model calculations, which gives some support
to the application of this perturbative description also for 
these small virtualities. 

\FIGURE[t]{
  \includegraphics[angle=270,  scale=0.8]{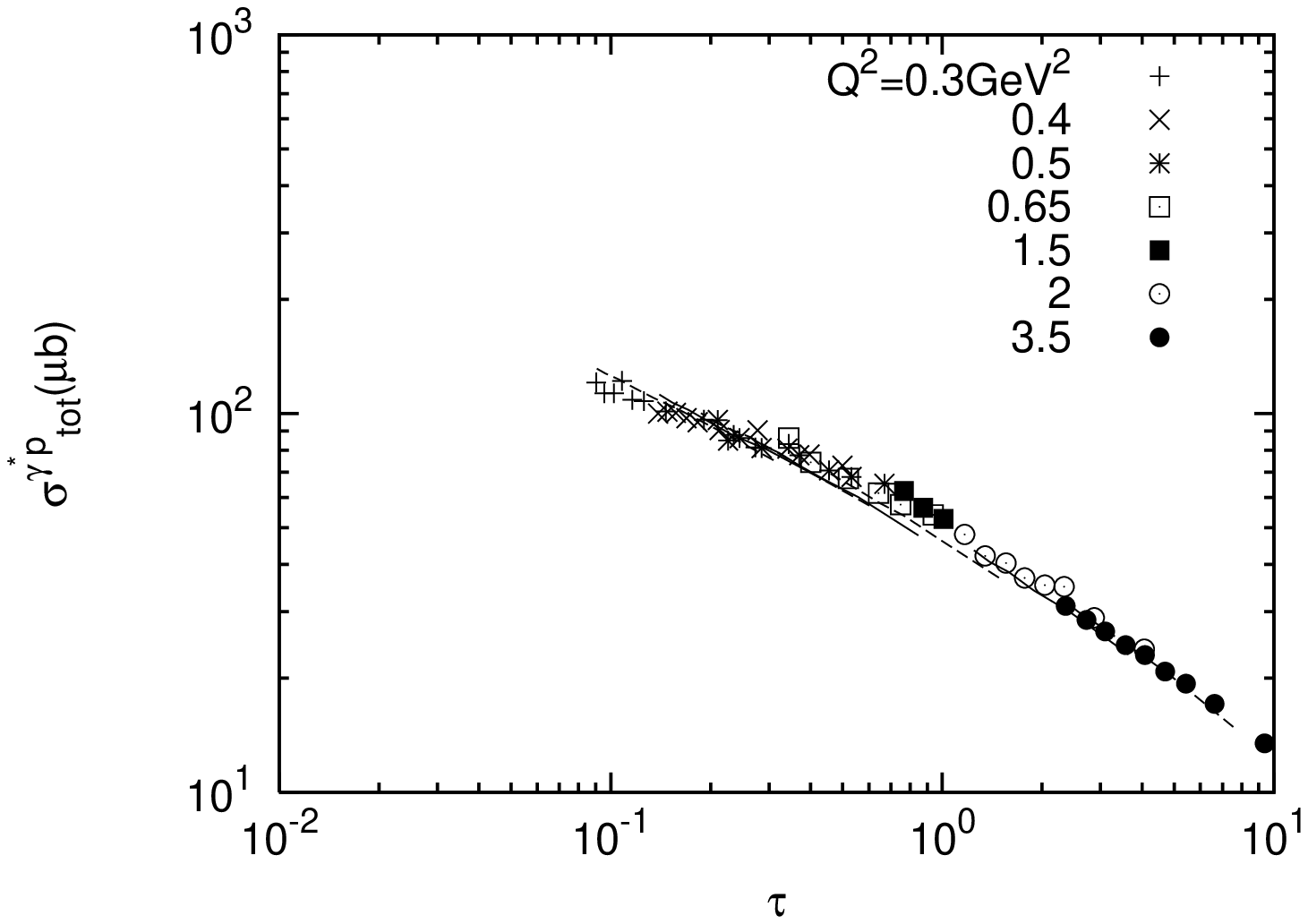}
  \caption{\label{fig:smallQ} The $\gamma^*p$ total cross section plotted as a 
    function of the scaling variable $\tau$ for low $Q^2$ values. Points are data from 
    the H1\cite{Adloff:2000qk} and ZEUS\cite{Breitweg:2000yn} collaborations, while the 
    lines are results obtained from our Monte Carlo. We can here see a successive 
    suppression of the cross section for smaller $Q^2$ as a result of the finite 
    quark masses.} 
}

The effective quark mass is quite small, and we would have expected
a larger suppression for small $Q^2$. The value 60 MeV is smaller than $\Lambda_{QCD}$,
and also smaller than the effective mass obtained in ref.~\cite{Dosch:1997nw}
in an analysis of the vector-current two-point function.
Ref.~\cite{Dosch:1997nw} studies a model with 
an effective quark mass which varies with $Q^2$, becoming smaller
when $Q^2$ is increased. The agreement in fig.~\ref{fig:smallQ} would
actually be improved by such a varying effective mass, but we do not believe 
that the accuracy of our model is sufficient to claim that
this improvement is significant. We have therefore here only used a constant 
effective quark mass in the wave functions in \eqref{eq:psigamma}.

\subsection{Is geometric scaling obeyed for $Q^2 / Q_s^2 < 1$?}

The effect of a finite quark mass is approximately
a multiplicative factor which suppresses the cross section for
smaller $Q^2$. As discussed in sec.~\ref{sec:charm} the result of the quark mass 
in the photon wave functions corresponds roughly to a suppression by
a factor $Q^2/(Q^2+4m_f^2)$.
The fact that
the model results in fig.~\ref{fig:smallQ} look as a scaling function is therefore
a consequence of the strong correlation between $x$ and $Q^2$ in the 
experimental data. The finite energy in the HERA accelerator
implies that for small values 
of $\tau$ there are only data for ($x$, $Q^2$)-values within a very 
small interval.
With a future accelerator with higher energy, one could also reach smaller 
$\tau$-values keeping $Q^2> 1\mathrm{GeV}^2$. For these $Q^2$ 
we do not get much suppression from the light quark masses, and fig.~\ref{fig:smallQ2}
shows that the scaling curve will lie somewhat above the present HERA results. 
In the figure we show the extend the curve for $Q^2=$2GeV$^2$ to $\tau\approx 0.07$.  
As can be seen, the difference is increasing for smaller $\tau$-values and is about a factor 1.4 
for $\tau \approx 0.07$. We also expect to see the scalebreaking effects of the charm 
mass for 1GeV$^2 \lesssim Q^2 \lesssim 10$GeV$^2$, while for higher $Q^2$ these 
effects should be gradually switched off. This cannot, however, be observed at 
accelerators in the foreseeable future.  

\FIGURE[t]{
  \includegraphics[angle=270,  scale=0.8]{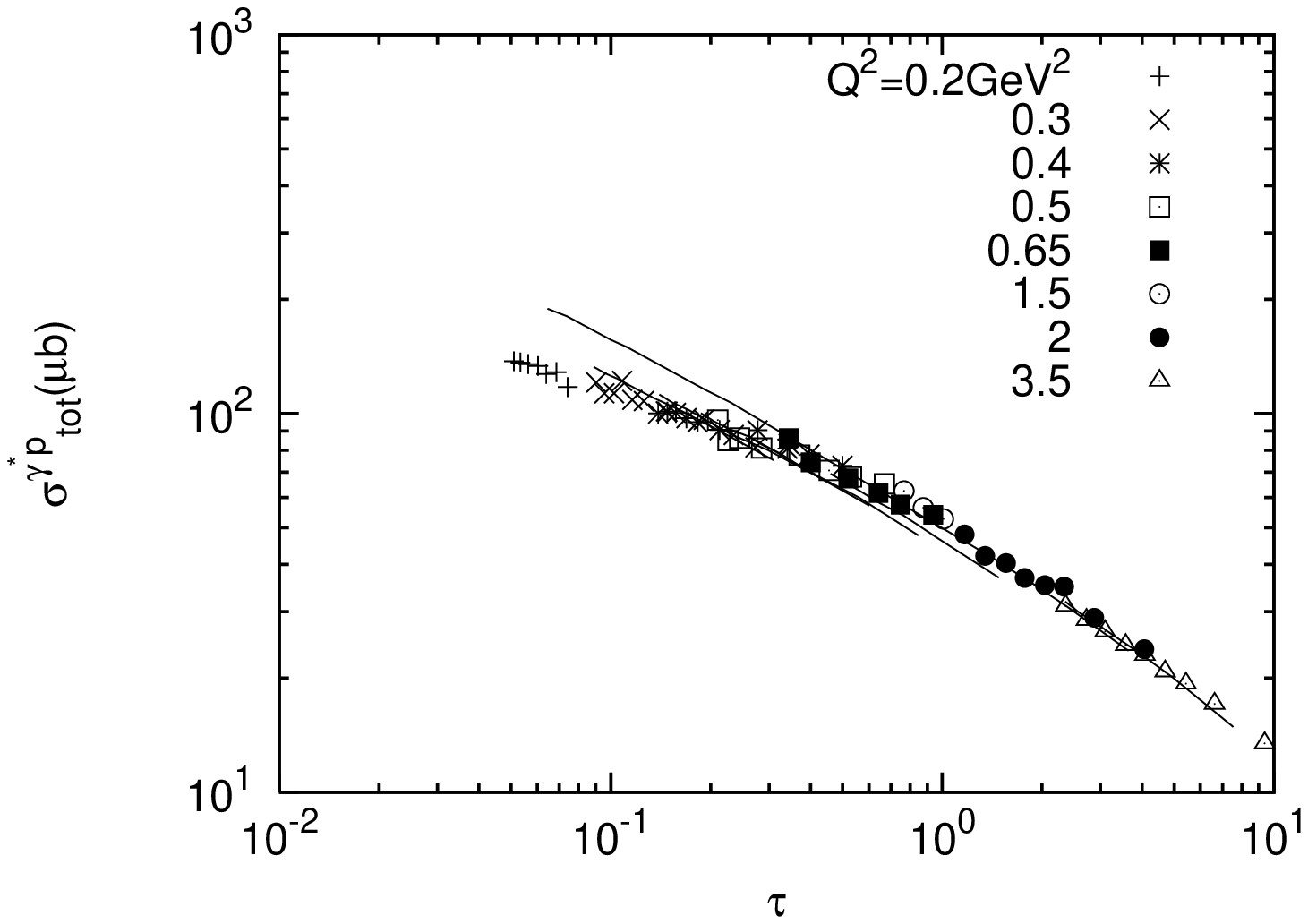}
  \caption{\label{fig:smallQ2} In this figure we extend the result
    for $Q^2 = 2$GeV$^2$ to smaller $\tau$-values. 
    We can here see a deviation from the scaling behaviour 
    at higher energies, about a factor 1.4 for $\tau \approx 0.07$
    corresponding to $x \approx 3\cdot 10^{-9}$ for $Q^2 = 2$GeV$^2$.
    In this plot the deviation from the scaling curve is 
    to a large extent due to the effective light quark mass. }  
}

As a conclusion of this section we predict that at higher energies the data will
scale at larger values for the cross section than the present HERA data for $\tau<1$. 
We should, however, point out that the analysis in this section only holds for the 
particular value $\lambda=0.3$ for the saturation scale in eq. \eqref{eq:R0}. 
For a different $\lambda$ the scaling behaviour will be different and it turns 
out that the deviation from the scaling behaviour would be 
somewhat reduced with a higher $\lambda$.

\section{Impact parameter dependence of the traveling wave}
\label{sec:impact}

In this section we will look at the impact parameter dependence of the 
dipole-proton scattering amplitude $T^{dp}(b)$, defined as $T=1-S$. 
We expect that the contribution from small dipoles is reduced
for large impact parameters,
which may result in different scaling behaviour for central and peripheral
collisions. This feature may be of special interest in studies of diffraction.

The amplitude $T^{dp}$ is
related to the total dipole-proton cross section by the relation
\begin{equation}
\sigma^{dp}(z,\pmb{r}) = 2\int d^2\pmb{b}\, T^{dp}(b).
\end{equation}
To get the amplitude for photon-proton scattering, the amplitude $T^{dp}(b)$
has to be weighted by the photon wave function in eq.~(\ref{eq:psigamma}).
The average value $\langle T\rangle$ will then depend on the virtuality 
$Q^2$ and the impact parameter $b$.

In the introduction we mentioned the similarity between the high energy 
evolution equations for $T$ and a certain type of equation (or rather a
class of equations) from statistical physics, known as the FKPP equation.  
Neglecting the impact parameter dependence, the BK equation for the
amplitude $T(l,Y)$ (with $l\equiv \mathrm{ln}k^2$)
is analogous to the FKPP equation for the function $u(x,t)$, with the 
identifications $l \leftrightarrow x$ and $\bar{\alpha}Y \leftrightarrow t$. 
The asymptotic solution, as $t \rightarrow \infty$, $u_{as}(x,t) = u(x-vt)$, then corresponds
to $T_{as}(l,Y) = T(l-\lambda Y)$ in QCD. If we define $Q_s^2=\mathrm{exp}(\lambda Y)$ as 
before the solution satisfies the geometric scaling relation 
$T(l,Y)_{as} = T(\mathrm{ln}k^2/Q_s^2)$. There are a number of conditions which must be satisfied 
in order to obtain an asymptotic solution of this form, and for a short 
review on traveling wave solutions in QCD we refer to \cite{Munier:2006zf}. 

\FIGURE[h]{
  \includegraphics[angle=270, scale=0.5]{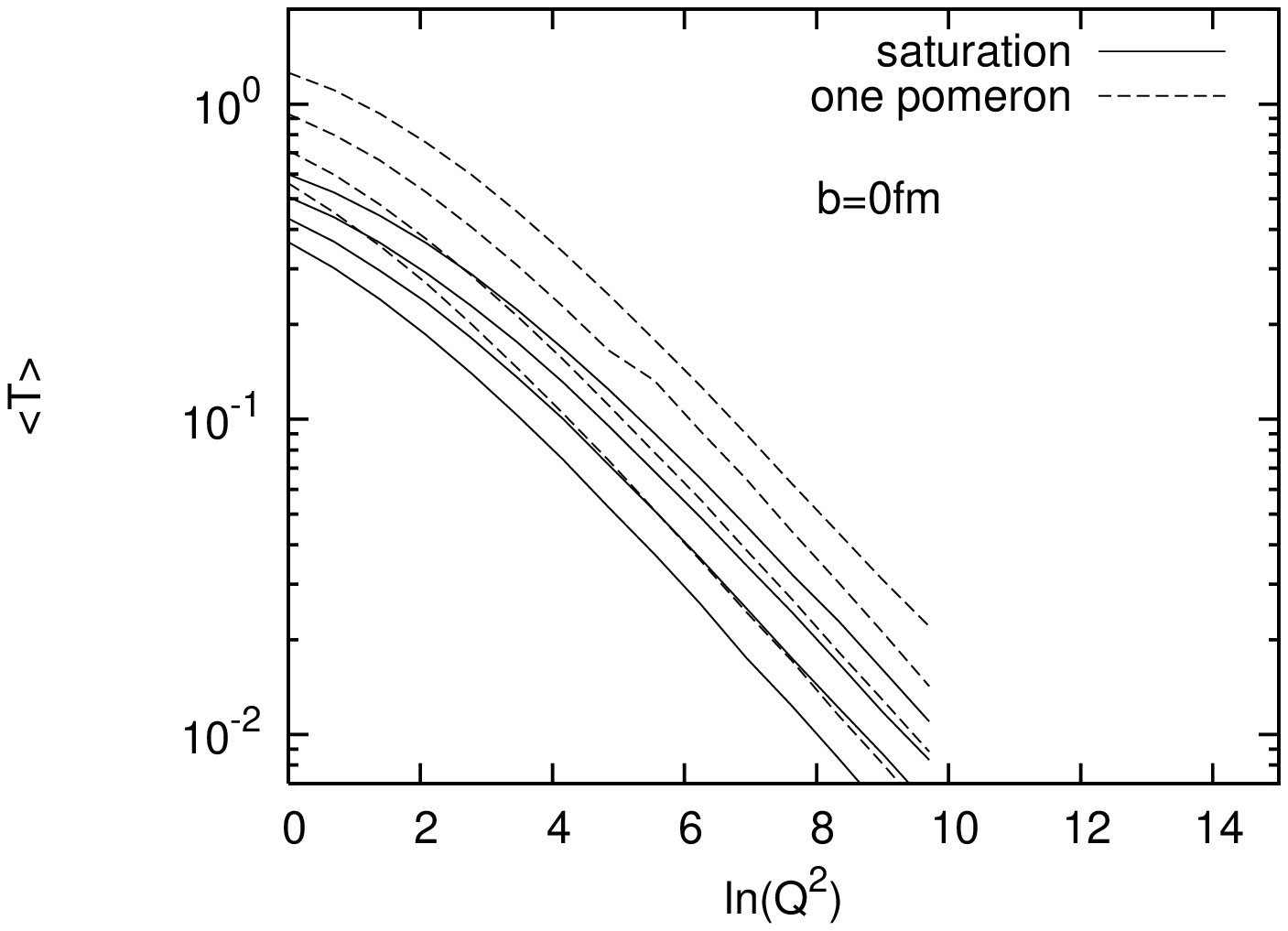}\includegraphics[angle=270, scale=0.5]{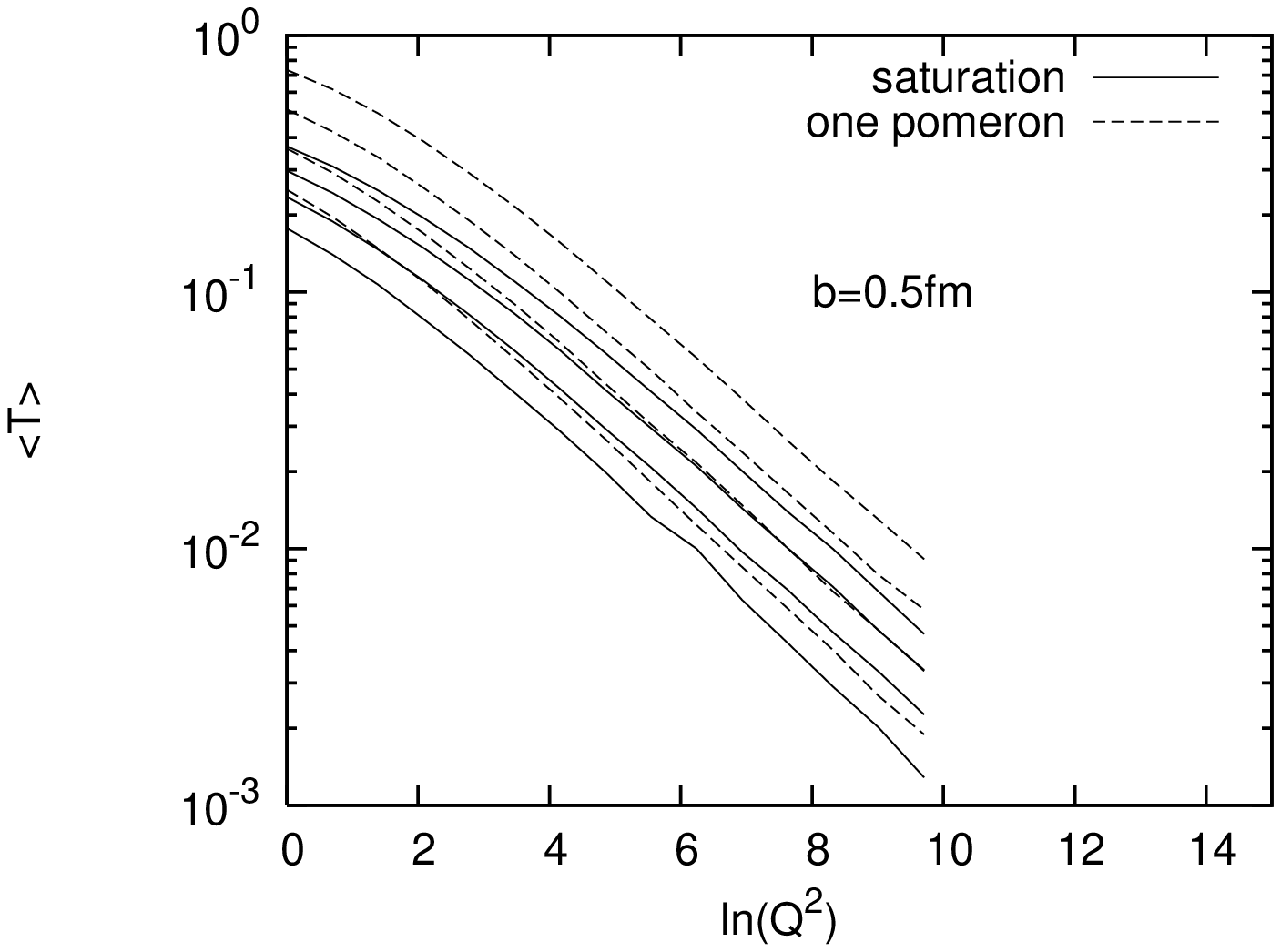}
  \includegraphics[angle=270, scale=0.5]{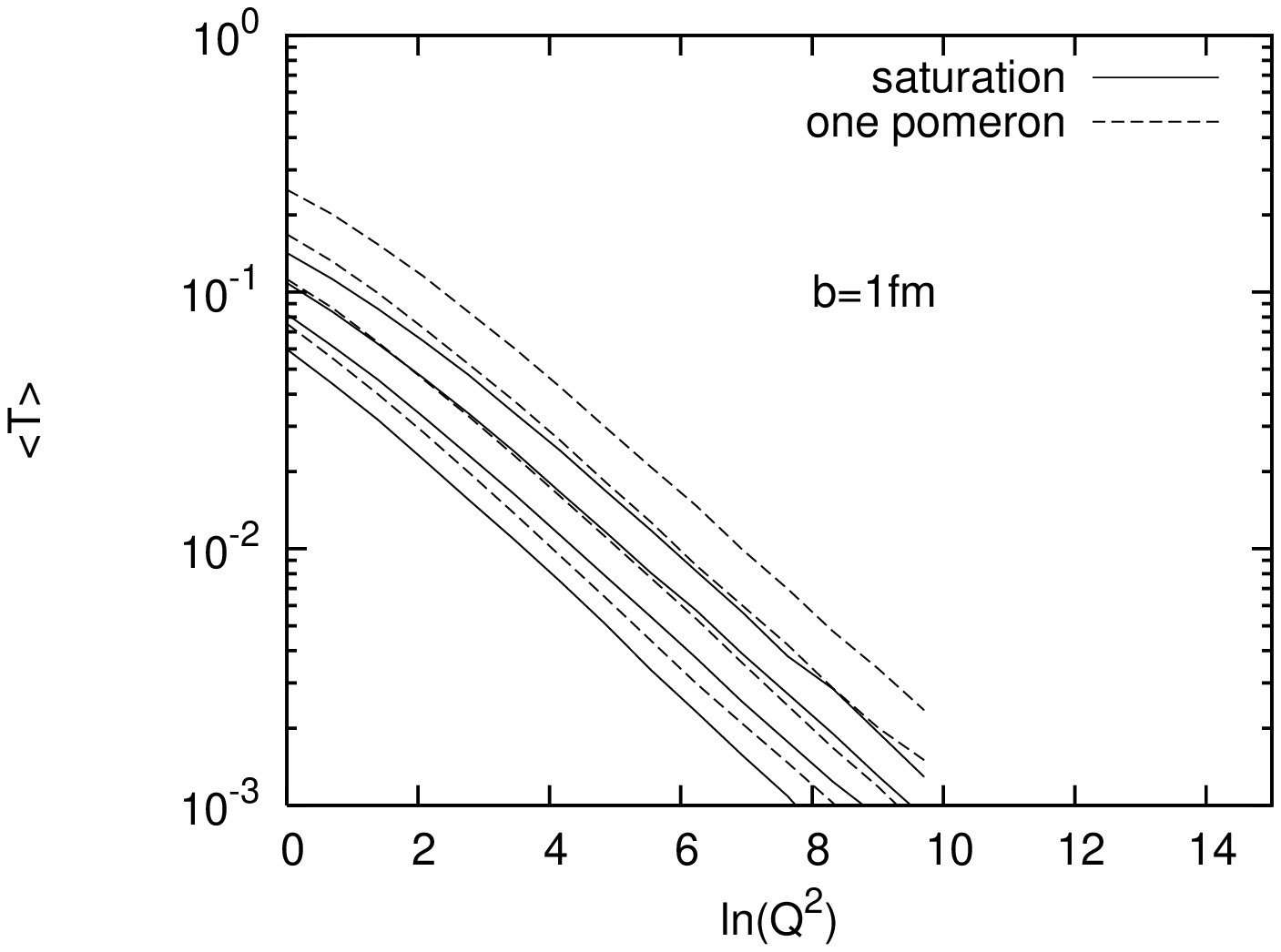}
  \caption {\label{fig:Tevolutionb} The average scattering amplitude as a function 
    of $\mathrm{ln}(Q^2/1\mathrm{GeV}^2)$ for $Y=6, 8, 10$ and $12$
    and impact parameters $b=0, 0.5, 1$fm. Both the amplitude containing 
    full saturation effects and the linear one pomeron 
    amplitude are shown. For the amplitude including saturation 
    we obtain from here the velocities 
    $v(b=0)=0.28$, $v(b=0.5fm)=0.35$ and $v(b=1fm)=0.37$. } }

As our model satisfies geometric scaling we can expect that the function 
$\langle T\rangle(\mathrm{ln} Q^2)$ plotted for different $Y$-values looks like a traveling wave. 
This can be seen in fig. \ref{fig:Tevolutionb} which shows results for
$Y=6,8,10$ and $12$. In this figure we also show the
result for different impact parameters $b$. 
If we look at different points with the same $\langle T\rangle$-value,
the velocity is determined by the relation
\begin{eqnarray}
v(b)=\frac{\Delta \mathrm{ln}Q^2}{\Delta Y}
\end{eqnarray} 
The results in fig.~\ref{fig:Tevolutionb} correspond to the following
velocities when saturation effects are included in the amplitude: 
$v(b=0)=0.28$, $v(b=0.5fm)=0.35$ and $v(b=1fm)=0.37$.
(For the one pomeron amplitude we instead obtain $v(b=0)=0.41$, 
$v(b=0.5fm)=0.46$ and $v(b=1fm)=0.48$) 
Thus the velocity varies significantly
with the impact parameter, being smaller for central collisions
and larger for peripheral collisions. As the scattering probability is largest 
for the central collisions, these results seem to be reasonably
consistent with weighted average over impact parameters corresponding to
the velocity $v \approx 0.3$ observed for the total cross section.

In fig.~\ref{fig:Tevolutionb} we can see that also the one pomeron amplitude, 
which contains no saturation effects, seems to exhibit the form of a traveling wave. 
We also see that we need quite high energies, corresponding to $x\approx 6\cdot 10^{-6}$ for 
$Q^2\approx 4 \mathrm{GeV}^2$ and $b=0$, 
before the one pomeron amplitude reaches the unitarity limit $T=1$. However, 
it is also seen that saturation effects reduce the amplitude by about $10\%$ for 
larger $Q^2$ and about $40\%$ for small $Q^2 \sim 1\mathrm{GeV}^2$ already 
at $x\sim 10^{-3}$ for $b = 0$. This result is consistent with the previous results 
presented in fig. \ref{fig:mcscaling2}.

\section{Conclusions}
\label{sec:conclusion}

DIS data from HERA show a striking regularity
as $\sigma^{\gamma^* p}$ is a function of the ratio $\tau=Q^2/Q_s^2(x)$ only, with
$Q_s^2(x)$ given by eq.~(\ref{eq:R0}) 
\cite{Stasto:2000er,Golec-Biernat:1998js,Golec-Biernat:1999qd}. Such a geometric scaling has 
been expected 
in the range $\tau<1$, as a natural consequence of saturation when the gluon 
density becomes very high, and there has been a lot of discussion 
in the literature whether saturation has or has not been observed at HERA.   
Modifications of the saturation model including
DGLAP evolution \cite{Bartels:2002cj} improves the agreement with data for larger $Q^2$,
but it has not been equally obvious how
geometric scaling follows in a natural way from QCD evolution in the non-saturated 
domain. 
The traveling wave solutions to the nonlinear evolution
equations in QCD do predict geometric scaling also for $\tau > 1$ 
\cite{Munier:2003vc, Munier:2004xu}, but these 
solutions are valid only at extremely high energies, far beyond what is available at 
the HERA accelerator.   

In this paper we have tried to shed some light on these questions: What
is the reason for scaling for $\tau>1$, and is saturation indeed the 
reason for geometric scaling for $\tau<1$? 
Supporting the saturation idea is the fact that the scaling curve seems to 
have a break just around $\tau=1$. There are, however, also other effects
which contribute to a break in this region. In ref.~\cite{Gustafson:2001iz} it 
is shown that the transition between $k_\perp$-ordered (DGLAP-like) chains and 
$k_\perp$-non-ordered (BFKL-like) chains is expected close to $Q_s^2$. Secondly
the finite energy in the HERA machine implies that experimental data for
$\tau<1$ are only available for $Q^2 < 1\mathrm{GeV}^2$, where nonperturbative
effects begin to be important.

To study the influence on the scaling from different dynamical effects
we have used the dipole cascade model presented in ref.~\cite{Avsar:2006jy}. This model
is based on Mueller's dipole cascade model, but includes energy-momentum
conservation and saturation effects, not only from multiple subcollisions
but also from pomeron mergings in the cascade evolution, and it does successfully reproduce
the data from HERA. 

Our conclusion is that scaling is a natural consequence of the dipole evolution
in the linear region $\tau>1$. Indeed, neglecting saturation the linear evolution
exhibits geometric scaling at high energies for $\tau$-values both larger and 
smaller than 1. The change from the DGLAP to the BFKL regime causes a change
in the slope of the scaling curve, but this change is rather smooth,
without a sharp break at $\tau=1$. The break seen in the HERA data is to a significant part 
caused by saturation, but to an even larger extent by scalebreaking effects at low $Q^2$.
The latter are partly due to the large $c$-quark mass and partly due to
non-perturbative effects related to 
confinement for the light $u$-, $d$-, and $s$-quarks. 
Therefore we predict that the results
from a future higher energy machine will show deviations from the scaling behaviour 
in the variable $\tau_{GW}$. These results are expected to lie above the HERA results 
for $\tau_{GW} < 1$, and be represented by a curve which is more smooth around $\tau_{GW}=1$.
(Recently it has also been predicted that a new type of scaling, called diffusive
scaling, will occur at very high energies, but we have in this paper concentrated 
on energies which might be within reach at some future accelerator.)

We study the charm contribution separately, and compare with data for the charm 
cross section. The charm contribution does not scale as the total cross
section, but we see that it does scale rather well as a function of the variable
$(Q^2 + 6m_c^2)/Q_s^2(\bar{x})$. Finally we have studied how the scaling feature
varies with impact parameter. We here see that the effective power $\lambda$ in
\eqref{eq:R0} is somewhat smaller for central collisions and larger for 
peripheral collisions, averaging out to the observed value around 0.3.
Such a variation may be of interest in studies of diffractive scattering.

\section*{Acknowledgments}

We want to thank Leif L\"onnblad for valuable discussions and 
help with the MC simulation. 
We are also grateful to Hannes Jung for
pointing out to us the lack of scaling in the charm contribution,
and to Edmond Iancu for useful comments on the manuscript. 

\bibliographystyle{utcaps}
\bibliography{/home/shakespeare/people/leif/personal/lib/tex/bib/references,refs}

\end{document}